\documentclass[pre,aps,reprint,superscriptaddress]{revtex4-1}
\usepackage{graphics}
\usepackage{graphicx}
\usepackage{amsmath}
\usepackage{float}
\usepackage{xcolor}
\usepackage{ulem}

\newcommand{\beginappendix}{%
        \setcounter{table}{0}
        \renewcommand{\thetable}{A\arabic{table}}%
        \setcounter{figure}{0}
        \renewcommand{\thefigure}{A\arabic{figure}}%
}

\begin{document}

\title{Criticality in sheared, disordered solids. I. Rate effects in stress and diffusion} 

\author{Joel T. Clemmer}
\affiliation{Sandia National Laboratories, Albuquerque, New Mexico 87123, USA}
\author{K. Michael Salerno}
\affiliation{Army Research Lab, Aberdeen, Maryland 21005, USA}
\author{Mark O. Robbins}
\affiliation{Department of Physics and Astronomy, Johns Hopkins University, Baltimore, Maryland 21218, USA}
\date{\today}

\begin{abstract}

Rate-effects in sheared disordered solids are studied using molecular dynamics simulations of binary Lennard-Jones glasses in two and three dimensions.
In the quasistatic (QS) regime, systems exhibit critical behavior: the magnitudes of avalanches are power-law distributed with a maximum cutoff that diverges with increasing system size $L$.
With increasing rate, systems move away from the critical yielding point and the average flow stress rises as a power of the strain rate with exponent $1/\beta$, the Herschel-Bulkley exponent. 
Finite-size scaling collapses of the stress are used to measure $\beta$ as well as the exponent $\nu$ which characterizes the divergence of the correlation length.
The stress and kinetic energy per particle experience fluctuations with strain that scale as $L^{-d/2}$.
As the largest avalanche in a system scales as $L^\alpha$, this implies $\alpha < d/2$.
The diffusion rate of particles diverges as a power of decreasing rate before saturating in the QS regime.
A scaling theory for the diffusion is derived using the QS avalanche rate distribution and generalized to the finite strain rate regime.
This theory is used to collapse curves for different system sizes and confirm $\beta/\nu$.

\end{abstract}

\maketitle

%=====================================================================%
\section{Introduction}
%=====================================================================%

Yield stress materials, or Bingham plastics, are substances that do not flow unless a critical yield stress $\sigma_c$ is exceeded \cite{Bonn2017}. 
This type of behavior has been identified in a wide variety of disordered systems including foams \cite{Park1994}, bubble rafts \cite{Durian1995, Dennin2004}, emulsions \cite{Mason1996}, colloids \cite{Coussot2002}, granular media \cite{Miller1996, Hayman2011}, and bulk metallic glasses \cite{Sun2010,Sun2012,Antonaglia2014}.
The athermal transition from a jammed state to an unjammed state at $\sigma_c$ is known as the yielding transition \cite{Lin2014} and is related to other unjamming transitions \cite{Liu2010}. 

The behavior of yield stress materials at this critical point is remarkably similar to the critical scaling at the onset of motion in other systems \cite{Fisher1998}. In particular, the yielding transition resembles the depinning transitions
of domain walls in magnets \cite{Ji1991}, fluid interfaces in porous media or on rough surfaces \cite{Martys1991}, and crack fronts moving through disordered solids \cite{Maloy2006}.
Before yielding, the system undergoes local plastic rearrangements or avalanches as the stress $\sigma$ is quasistatically increased towards $\sigma_c$.
Above $\sigma_c$, the system flows with an average strain rate $\dot{\epsilon}$ that scales as a power of the excess stress: $\dot{\epsilon} \sim (\sigma -\sigma_c)^\beta$. Here $\beta$ is a critical exponent whose inverse $n=1/\beta$ is commonly known as the Herschel-Bulkley exponent \cite{Herschel1926}.
Plastic activity is correlated over a length that diverges at the yield stress: $\xi \sim |\sigma - \sigma_c|^{-\nu}$.
As $\sigma$ approaches $\sigma_c$ from below,
$\xi$ is related to the divergence of the linear extent of the largest
avalanches.
As $\sigma$ approaches $\sigma_c$ from above, the growing correlation length reflects larger spatial and temporal fluctuations in the strain rate
that can be related to the avalanches below $\sigma_c$.

In the thermodynamic limit, the yielding transition corresponds to an infinitely small strain rate. 
In a finite system, which is typically studied due to experimental or computational constraints, the transition from jamming below $\sigma_c$ to flowing above $\sigma_c$ is broadened over a range of stresses. 
Therefore, the yielding transition is often studied using constant strain rate as opposed to constant values of the stress. 
Experiments on bulk metallic glasses and granular packings \cite{Sun2010, Sun2012, Antonaglia2014, Tong2016, Denisov2016, Bares2017} as well as molecular dynamics (MD) simulations \cite{Salerno2012, Salerno2013} 
at quasistatic strain rates have identified a power-law distribution of the magnitude of avalanches. 
The size of the largest avalanche has been found to diverge with increasing system size \cite{Salerno2012, Salerno2013}, and the critical exponents characterizing their size were determined from MD simulations using finite-size scaling. 

Elastoplastic models (EPM) have also been used to advance theoretical understanding of the yielding transition \cite{Nicolas2018}.
In EPMs, solids are coarse-grained to a lattice and each cell tracks the local evolution of stress.
When the local stress exceeds a threshold, a cell will plastically yield and redistribute stress to neighboring cells.
Studies of quasistatically driven EPMs have found similar avalanche statistics \cite{Talamali2011, Budrikis2013, Lin2014, Lin2014a, Liu2016, Budrikis2017, Karimi2017, Ferrero2019, Tyukodi2019}.

It has been suggested that yielding may be in the same universality class as interface depinning, earthquakes, and other systems \cite{Dahmen2011, Salje2014}, but simulations have shown the critical exponents are different \cite{Salerno2012, Salerno2013, Lin2014}.
The biggest discrepancy is in the rate of avalanche nucleation, which scales extensively with system size for depinning \cite{Martys1991a,Clemmer2019} and subextensively for yielding \cite{Salerno2012, Salerno2013,Tyukodi2019}. 
Lin, Lerner, Rosso, and Wyart have argued that the yielding transition is distinct from the depinning transition due to the nature of the elastic interactions in the two types of system \cite{Lin2014}.
In depinning, interactions always have the same sign. 
If one region of the interface advances, it pulls all nearby regions with it. 
In contrast, the activation of a shear transformation zone, the fundamental unit of rearrangement in a sheared disordered solid \cite{Falk1997}, produces a quadrapolar stress field \cite{Langer2001}. 
Thus as one region of the solid relaxes it may either stabilize or destabilize neighboring regions depending on their relative position. 
This variation in the sign of the elastic coupling means that sheared systems do not obey the ``no-passing rule'' \cite{Middleton1992, Middleton1993} that constrains interface depinning models.

In this work we study the limit of finite strain rates (FSR) and the transition to quasistatic (QS) shear using MD simulations of 2D and 3D disordered solids containing up to $7 \times 10^6$ particles. 
We use finite-size scaling techniques to accurately measure several critical exponents in 2D and 3D, including $\nu$ and $\beta$.
We also propose new scaling relations that provide bounds on $\beta/\nu$ using measures of quiescence in the system. 

In Sec. \ref{sec:yield_methods}, we describe the simulation methods and initial system preparation. 
In Sec. \ref{sec:qs_scaling}, the scaling of QS avalanches is reviewed.
Based on the QS theory, a transition to the FSR regime is then described in Sec. \ref{sec:transition} and limits are placed on the strain rate of the transition. 
This transition is then tested in Sec. \ref{sec:steady} where the rate dependence of the average flow stress is collapsed using a finite-size scaling ansatz providing accurate measurements of $\sigma_c$, $\nu$, and $\beta$. 
The fluctuations in stress are then addressed in Sec. \ref{sec:yield_deviate}. 
In Sec. \ref{sec:yield_diffusion} we look at the diffusion of particles highlighting the importance of boundary conditions.
Finally in Sec. \ref{sec:yield_summary}, we conclude with a summary of our findings and a comparison to other results in MD and EPM.

%%%%%%%%%%%%%%%%%%%%%%%%%%%%%%%%%%%%%%%%%%%%%%%%%%%%%%%%%%%%
\section{Methods}
\label{sec:yield_methods}
%%%%%%%%%%%%%%%%%%%%%%%%%%%%%%%%%%%%%%%%%%%%%%%%%%%%%%%%%%%%

We simulate pure shear of two and three dimensional disordered packings using MD.
The systems are bidisperse and are similar to models used in other work studying the yielding transition \cite{Maloney2006, Maloney2008, Maloney2009, Salerno2012, Salerno2013}.
The two types of disks or spheres are labeled $A$ and $B$ and have the same mass $m$.
Particles of type $I$ and $J$ through an attractive Lennard-Jones (LJ) potential:
\begin{equation}
U_{IJ}(r) = 4 u_{IJ} \left[(a_{IJ}/r)^{12}-(a_{IJ}/r)^6\right] \ \ ,
\end{equation}
where $r$ is the distance between the two particles, $a_{IJ}$ is a diameter, and $u_{IJ}$ is an interaction strength.
To limit the range of interactions, the potential is smoothly interpolated to zero at $r_c = 1.5 a_{IJ}$. This is accomplished using a fourth order polynomial function that starts at a distance of $1.2 a_{IJ}$.

Particles of type $A$ and $B$ have radii $0.5 a$ and $0.3 a$, respectively,
where $a$ is taken as the unit of length.
The radii are additive, so the effective diameters are
$a_{AA}=a$, $a_{AB}=0.8a$, and $a_{BB} = 0.6 a$.
The self-interaction strengths are $u_{AA} = u_{BB} = u$, where $u$ is taken as the fundamental unit of energy.
We have considered two values of the cross interaction, $u_{AB} = u$ and $2u$. 
Increasing the strength of the cross interaction encourages the system to mix so we refer to the two choices as the neutral and mixing models, respectively.

The values of $a_{IJ}$ are chosen to help ensure that the shearing system remains disordered by adding geometrical frustration \cite{Lan1988, Maloney2006}.
While this bidispersity increases the free energy barrier to nucleate and grow crystalline domains \cite{VanMeel2009},
the ground state of the neutral model is still a phase separated, crystalline state. 
In 2D, phase segregation was observed at large strains.
In contrast, the ground state of the mixing model is expected to be a mixed configuration due to energetically favorable cross interactions.
No evidence of segregation or crystallization was observed in 
simulations of the mixing model and it is used for all results unless otherwise noted.

The fundamental unit of time is defined as $t_{0} = \sqrt{a^2 m/u}$.
All quantities in the following text are presented in units of $a$, $u$, $t_0$, or appropriate combinations. For example strain rates are in units of $t_0^{-1}$ and
stress is in units of $u/a^d$, where $d$ is the spatial dimension.
Simulations were run in LAMMPS using the velocity-Verlet algorithm with a timestep of $\Delta t = 0.005 $ \cite{Plimpton1995}.

Initial particle configurations were prepared in a manner similar to other works \cite{Maloney2006, Maloney2008}. Particles were randomly placed in a square or cubic box 
with periodic boundary conditions and initial density $\rho_i$. 
The number of particles of type $N_A$ and $N_B$ had a fixed ratio of $N_A/N_B \approx (1+\sqrt{5})/4$.
A cosine potential was then applied between particles to separate overlapping particles for a time of about $25 $. 
This potential was then replaced with the LJ potential and the volume was changed over another time interval of $25$ to achieve
the desired final density $\rho$.
Simulations in 3D used $\rho_i= 1.8$ and $\rho = 1.7$ and the final cubic box length was $L=20.35$, 40.71, 81.42, or 162.83.
Simulations in 2D used $\rho_i= 1.6$ and $\rho= 1.4$, and the final square box had $L = 54.79$, 109.58, 219.16, 438.32, 876.64, or 1753.28.
These sizes are rounded to the nearest integer for the remainder of the paper.

The focus here is on steady-state shear of athermal, overdamped systems.
Systems were deformed under pure shear by applying an affine transformation to particle positions at a constant uniaxial strain rate $\dot{\epsilon}$.
As described below, the periodic cell is expanded along the $x$ direction and contracted in the other direction(s) to maintain constant volume.
The components of the stress tensor $\sigma_{\alpha \beta}$ were calculated from the virial and the kinetic energy $K$ associated with non-affine particle velocities $\vec{v}_{i,\mathrm{na}}$ which reflect deviations from the local environment \cite{Allen1989}.
The shear stress is defined as $\sigma \equiv (\sigma_{xx} - \sigma_{yy})/2$ in 2D and $\sigma \equiv (2\sigma_{xx} - \sigma_{yy} - \sigma_{zz})/4$ in 3D.

Work is done on the system at an average rate $L^d \dot{\epsilon} \langle \sigma \rangle$, where $\langle \sigma \rangle$ is the mean shear stress.
To maintain a steady state, energy was removed from the system using
a viscous damping force commonly used in Langevin thermostats. No Langevin noise term was added since the effective temperature is zero.
The damping force applied to particle $i$ is
$\vec{F}_{i,\mathrm{damp}} = - \frac{1}{2} \Gamma m \vec{v}_{i,\mathrm{na}}$.
The rate of work done by the thermostat on a particle is $-\frac{1}{2} \Gamma m \vec{v}_{i,\mathrm{na}} \cdot \vec{v}_{i}$.
The affine contribution to this work averages to zero, and the total dissipated power can be written as:
\begin{equation}
	P=\sum_i \frac{1}{2} \Gamma m \vec{v}_{i,\mathrm{na}}^2 =\Gamma \langle K \rangle = L^d \langle \sigma \rangle \dot{\epsilon} \ \ , 
	\label{eq:power}
\end{equation}
where angle brackets indicate a time average.
The damping coefficient is set to $\Gamma = 4$, which is well within the regime associated with the overdamped universality class of yielding in Refs. \cite{Salerno2012,Salerno2013}. 

Initial simulations used conventional periodic cells with fixed orientation.
The $x$ dimension of the periodic cell, $L_x$, was expanded at strain rate
$\dot{\epsilon} \equiv \frac{1}{L_x}\frac{dL_x}{dt}$, while the remaining dimensions were contracted to preserve area or volume.
In 2D, the $y$ dimension was contracted at $\dot{\epsilon}$, while for 3D, both the $y$ and $z$ dimensions were contracted at $\dot{\epsilon}/2$.
This contraction limits the maximum strain that can be applied because $L_y$ eventually becomes comparable to the range of interactions.
While we saw behavior characteristic of critical scaling, it became clear that a different approach was needed to attain steady state.

To access larger strains we imposed
the same pure strain deformation using Kraynik-Reinelt (KR) boundary conditions in 2D \cite{Kraynik1992} and generalized KR (GKR) boundary conditions in 3D \cite{Hunt2016}.
These methods deform the box shape and change the choice of periodic lattice vectors in a sequence of steps that prevents any cell dimension from becoming too small. 
Our implementation of these boundary conditions was heavily based on the source code of Nicholson and Rutledge \cite{Nicholson2016}. 
Modifications were made to apply the strain through an affine shift in particle positions as opposed to using the SLLOD equations of motion \cite{Evans1984}.
Energy introduced through shear is removed
by the viscous damping described above.

\begin{figure}
	\begin{center}
	\includegraphics[width=0.45\textwidth]{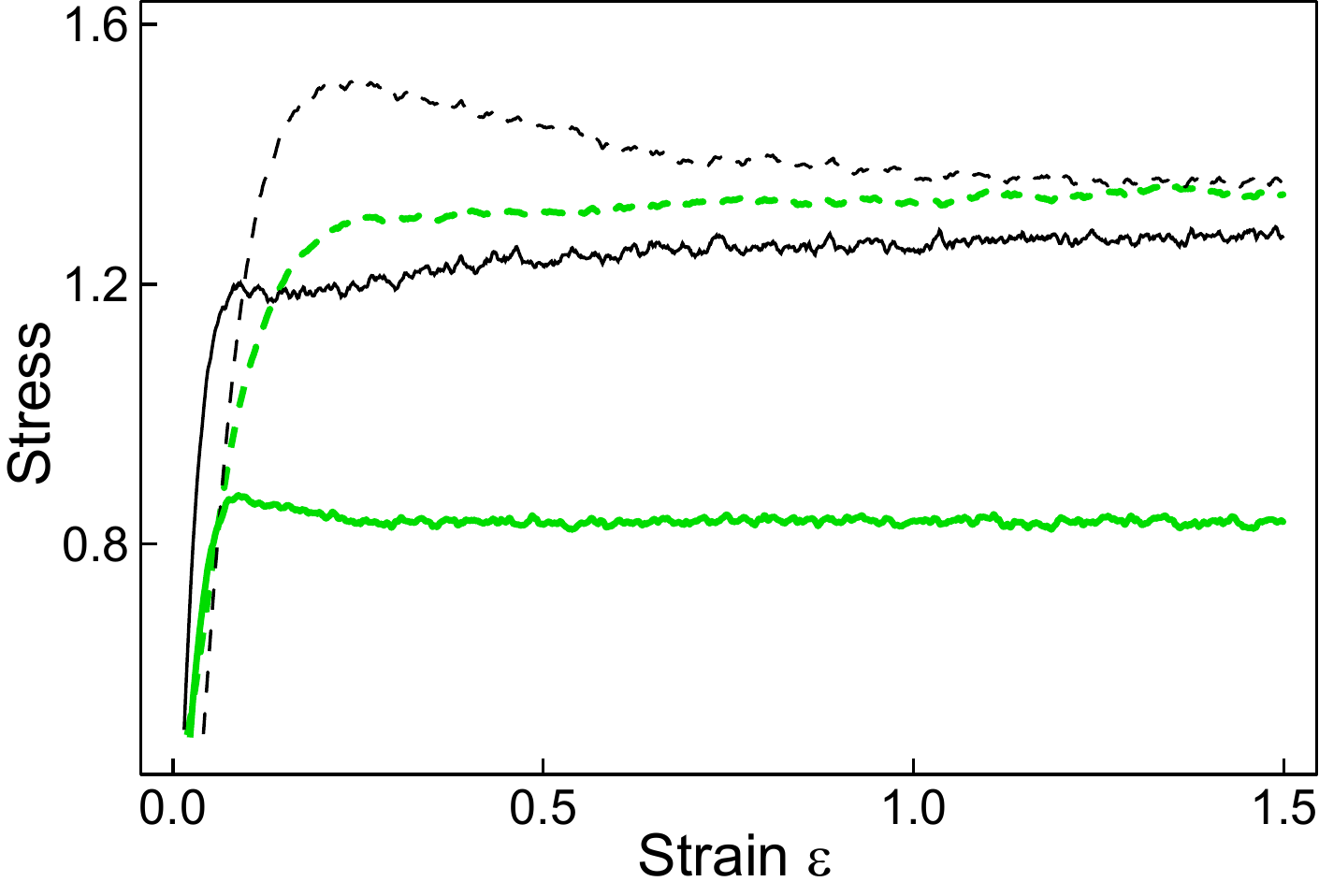}
	\caption{The shear stress (solid lines) and the pressure (dashed lines) are plotted as a function of strain for neutral (thick green) and mixing (thin black) models. Data is collected from a 3D system of size $L = 81$ at a rate of $\dot{\epsilon} = 2 \times 10^{-4}$.} \label{transitional}
	\end{center}
\end{figure}

Figure \ref{transitional} illustrates the evolution of shear stress 
$\sigma$ and 
pressure $p \equiv -(\sigma_{xx}+\sigma_{yy} + \sigma_{zz})/3$
with strain for the neutral and mixing potential.
This data is for 3D systems with $L=81$ and $\dot{\epsilon}= 2 \times 10^{-4} $, but similar results are seen for other systems and in prior work
\cite{Salerno2012,Salerno2013}.
For both potentials, there is a peak in $\sigma$ at about 7\% that indicates yield.
This initial yield stress is known to depend on the preparation of the initial state \cite{Varnik2004, Shi2005, Rottler2005, Ozawa2018}.

From Fig. \ref{transitional} we see that the shear stress shows a clear evolution with strain up to $\epsilon \sim 0.25$ or 0.5, while the pressure continues to evolve until strains of $1$ or more.
Studies of the radial distribution functions show that the structure evolves during this initial period.
The number of $AB$ neighbors increases for mixing interactions and decreases for the neutral potential, leading to segregated regions in 2D.

To ensure data is collected only after 
all memory of the initial preparation has been erased, 2D (3D) systems were first sheared to a strain of 50 (100) at a high rate of $\dot{\epsilon} = 10^{-3} $ ($2\times 10^{-3}$).
The strain rate was then slowly incremented downward, straining at each lower rate until a new steady state was reached.
At lower rates, simulations were run for longer times up to $\sim 10^6$, or $\sim$200 million timesteps, at the lowest rates.
We checked that simulations starting from lower rates and higher rates gave the same results.
Appendix \ref{sec:yield_simple} shows that simulations with simple shear are consistent with measured scaling exponents.

The applied strain rate $\dot{\epsilon} $ varied between $10^{-3}$ and $ 10^{-7}$ in 2D and $2 \times 10^{-3}$ and $2\times 10^{-7}$ in 3D.
During deformation we evaluated
the shear stress, pressure, kinetic energy, and diffusive motion of particles relative to the affine deformation,
$\langle \Delta \vec{r}^2 \rangle$.
These simulations were also used to derive the results in the companion paper, Ref. \cite{Clemmer2021}.

%%%%%%%%%%%%%%%%%%%%%%%%%%%%%%%%%%%%%%%%%%%%%%%%%%%%%%%%%%%%
\section{Avalanches and Quasistatic Flow}
\label{sec:qs_scaling}
%%%%%%%%%%%%%%%%%%%%%%%%%%%%%%%%%%%%%%%%%%%%%%%%%%%%%%%%%%%%

\begin{figure}
\begin{center}
	\includegraphics[width=0.45\textwidth]{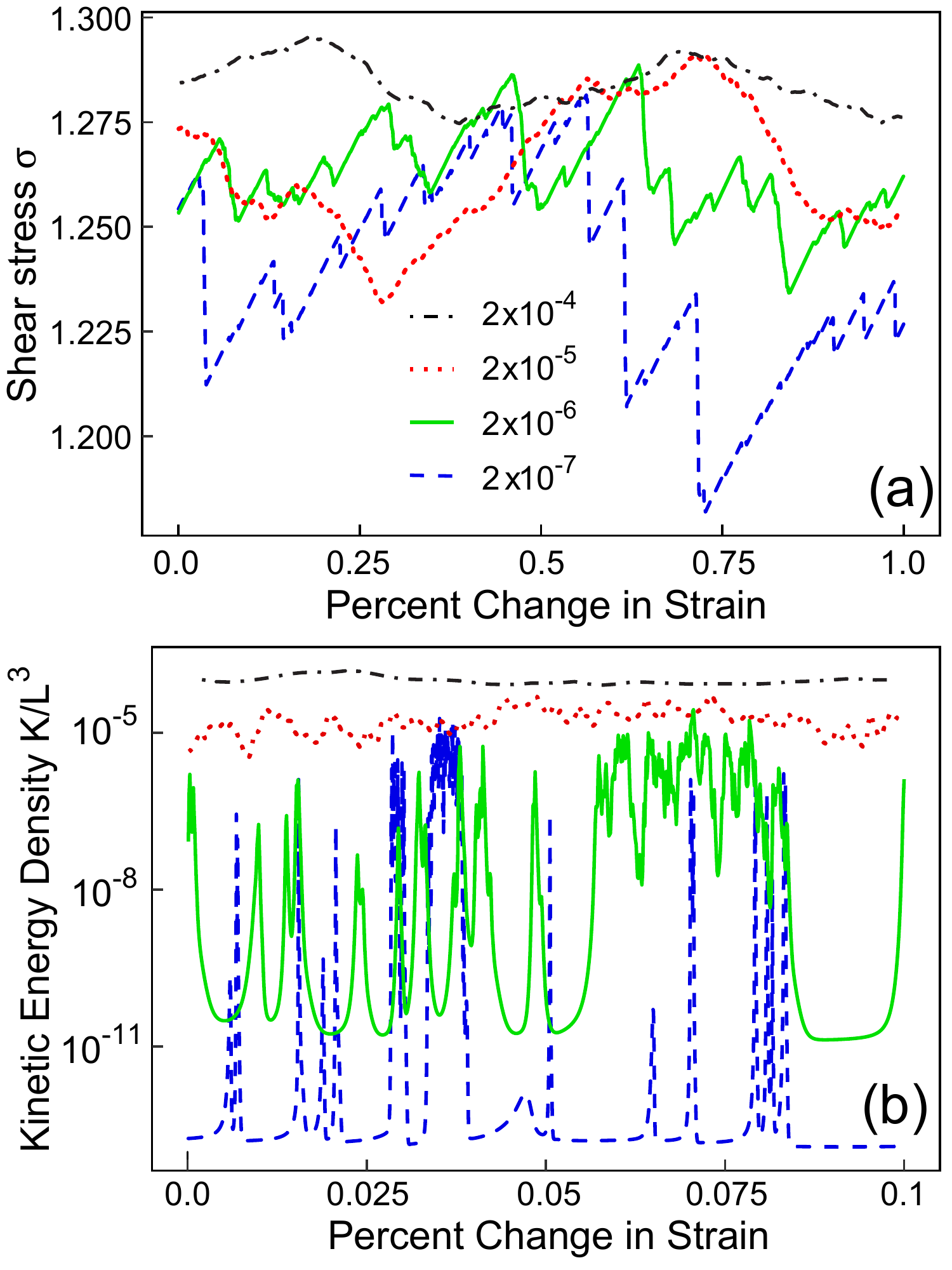}
	\caption{Example traces of (a) shear stress and (b) kinetic energy per particle, $K/L^3$, as a function of strain for a 3D system of size $L = 40$ in steady state. The system was strained at the rates indicated in the legend of (a). Note that the range of strains in (b) is smaller to reveal all avalanches. Small avalanches produce stress drops smaller than the line width in (a).
	The effective instantaneous temperature is of order $K/L^3$ and is always extremely small.
	} \label{steady_flow}
	\end{center}
\end{figure}

Figure \ref{steady_flow}(a) shows the variation of shear stress with strain in steady-state flow at the indicated strain rates.
The system is 3D with $L=40$, but similar trends are seen for all systems.
In this section, we discuss the low rate or quasistatic (QS) regime, illustrated by the results for $\dot{\epsilon} =2 \times 10^{-7}$.
The stress rises linearly as elastic energy is stored in the system
and then drops when the system becomes mechanically unstable, causing an avalanche of plastic rearrangement.
During each avalanche, stored elastic energy is converted into kinetic energy
as shown in Fig. \ref{steady_flow}(b). 
Since we are in the overdamped limit, the kinetic energy $K$ is proportional to
the rate of energy dissipation through plastic deformation.
During each avalanche, $K$ rises as plastic deformation spreads,
and then decays as the rate of plasticity drops back towards zero.
Note that the effective temperature associated with the kinetic energy per particle remains very low compared to the binding energy between particles, as required for the athermal limit.

In the QS regime, avalanches are well-separated in time and the kinetic energy drops down to a small background level before the next instability is triggered.
The background level $K_{b}$ is related to nonaffine displacements produced by the heterogeneity in elastic properties \cite{Maloney2015}. The corresponding nonaffine velocities scale as $\dot{\epsilon}$ and thus $K_{b} \propto \dot{\epsilon}^2$. It is negligible compared to the energy dissipated in avalanches which scales as $\dot{\epsilon}$ in the QS regime. 

One can divide the total kinetic energy released in avalanches into a sum over contributions $k_I$ from each avalanche $I$ that occur sequentially:
\begin{equation}
	K(t) = \sum_I k_I(t) \ \ .
	\label{eq:Kaval}
\end{equation}
Each avalanche dissipates an amount of energy
\begin{equation}
	E_I = \Gamma \int dt k_I(t) \\
	\label{eq:E_I}
\end{equation}
that is removed by the damping force.
As discussed in Ref. \cite{Salerno2012}, the magnitude of the associated drop in stress $\delta \sigma_I$ is proportional to $E_I$ for sufficiently large avalanches:
$E_I = L^d \langle \sigma \rangle \delta \sigma_I / 4 \mu$, where $\langle \sigma \rangle$ is the average stress and $\mu$ the shear modulus.
In general there is a correspondence between $K$ and $L^d d \sigma/dt$.
This connection is further expanded upon in the sibling paper \cite{Clemmer2021}.

The QS curves in Fig. \ref{steady_flow} show a broad distribution in the magnitudes of the stress drops $\delta \sigma_I$ and released energies $E_I$ during events.
The full range of variation is difficult to see in Fig. \ref{steady_flow}(a),
because the percentage change in $\sigma$ for small avalanches is negligible compared to the line width.
Many more small avalanches are evident in Fig. \ref{steady_flow}(b), which shows an expanded view of the
first 10\% of the strain interval in Fig. \ref{steady_flow}(a). 
The peak kinetic energy varies by more than 6 orders of magnitude for the avalanches shown and the integrated
energy dissipated varies even more.

Past studies of this system examined the critical scaling in the QS regime \cite{Salerno2012,Salerno2013}.
Both $E_I$ and $\delta \sigma_I$ follow a power-law distribution with the same exponent $\tau$ in the thermodynamic limit ($L \rightarrow \infty$).
The rate of avalanches of energy $E$ per unit strain is given by:
\begin{equation}
	R_{QS}(E,L) \sim L^\gamma E^{-\tau}
\label{eq:RQS}
\end{equation}
up to a maximum avalanche size $E_\mathrm{max} \sim  L^\alpha$, where $\alpha$ is also commonly denoted as $d_f$ in the literature as the fractal dimension.
The values of $\tau$, $\alpha$, and $\gamma$ are given in Table \ref{table:yield_exponents}.
As noted in the introduction, one of the surprising features is that $\gamma < d$ so that the rate of small avalanches grows more slowly than the size of the system.
This has also been seen in some EPMs \cite{Tyukodi2019}.

In the QS regime, one could theoretically expect the same sequence of avalanches occurs in a system independent of rate.
Increasing $\dot{\epsilon}$ just decreases the quiescent periods between avalanches (Fig. \ref{steady_flow}).
Several limiting results can be derived using this fact and the form of $R_{QS}$.
For example, $\langle K \rangle/\dot{\epsilon}$ is proportional to
$\langle \sigma \rangle$ (Eq. \eqref{eq:power}) and both become independent of rate at small $\dot{\epsilon}$.
One can write
\begin{equation}
	\langle K \rangle = \frac{1}{T} \int_0^T dt K(t)= \frac{1}{T} \sum_i \frac{E_i}{\Gamma} \ \ . 
\label{eq:meanK}
\end{equation}
The sum over avalanches can then be replaced by an integral over energies using the rate of avalanches per unit time $\dot{\epsilon} R_{QS}$.
One finds:
\begin{align}
\begin{split}
	\langle K \rangle &= \int dE \  \dot{\epsilon} \  R_{QS}(E,L) \  \frac{E}{\Gamma} \\
	&\sim \frac{\dot{\epsilon}}{\Gamma} L^\gamma \int^{E_\mathrm{max}} dE E^{1-\tau} \sim \dot{\epsilon} L^{\gamma+(2-\tau) \alpha} \ \ . 
\label{eq:meanK2}
\end{split}
\end{align}
From Eq. \eqref{eq:power}, $\langle K \rangle$ is proportional to the rate of power dissipated and scales as $\dot{\epsilon} L^d$.
This imposes the scaling relation 
\begin{equation}
	\gamma+(2-\tau)\alpha=d
	\label{eq:scale}
\end{equation}
found in Refs. \cite{Salerno2012,Salerno2013}.

%%%%%%%%%%%%%%%%%%%%%%%%%%%%%%%%%%%%%%%%%%%%%%%%%%%%%%%%%%%%
\section{Transition to a Finite Strain Rate Regime}
\label{sec:transition}
%%%%%%%%%%%%%%%%%%%%%%%%%%%%%%%%%%%%%%%%%%%%%%%%%%%%%%%%%%%%

In the limit of infinite system size, the system will be jammed ($\dot{\epsilon}=0$) if a constant stress less than $\sigma_c$ is applied.
At $\sigma > \sigma_c$, the system will flow at a finite rate that grows with the distance to the critical stress:
\begin{equation}
\dot{\epsilon} \sim (\sigma -\sigma_c)^\beta \ \ ,
\label{eq:HB}
\end{equation}
where $\beta$ is a critical exponent. This power-law scaling is commonly known as the Herschel-Bulkley law, with the Herschel-Bulkley exponent $n=1/\beta$ \cite{Herschel1926}. At stresses sufficiently close to $\sigma_c$, one expects to see critical behavior and a unique value of $\beta$ for a wide class of materials that all reside in the same universality class.
Our simulations are at constant strain rate, but Eq. \eqref{eq:HB} still applies
in the thermodynamic limit.

Figure \ref{steady_flow} shows an increase in the mean stress with increasing shear rate that is qualitatively consistent with Eq. \eqref{eq:HB} 
\footnote{Note however that snapshots of $\sigma$ over short strain intervals do not have a clear trend with rate because of
the large fluctuations in the instantaneous shear stress discussed in Sec. \ref{sec:yield_deviate}}.
In the QS regime, each avalanche has time to evolve. There is little or no change in $\langle \sigma \rangle$ with rate and Eq. \eqref{eq:HB} does not apply due to the finite size of the system.
As the strain rate increases, new mechanical instabilities are nucleated
before the previous avalanche finishes.
Both the stress and kinetic energy become smoother with increasing rate as more avalanches overlap in time and/or space.
In Fig. \ref{steady_flow}, increasing $\dot{\epsilon}$ from $2 \times 10^{-7}$ to $2 \times 10^{-6}$ reduces the maximum size of stress drops and $K$ does not decrease as significantly between some stress drops.
By $\dot{\epsilon} = 2 \times 10^{-5}$, the stress shows undulations rather than sharp drops and
one can no longer distinguish individual avalanches or quiescent periods between avalanches in the kinetic energy.
This transition occurs at lower rates as $L$ increases. 
For an infinite system, this transition goes to a strain rate of zero such that Eq. \eqref{eq:HB} is valid at arbitrarily small rates.

\begin{figure*}
\begin{center}
	\includegraphics[width=1.0\textwidth]{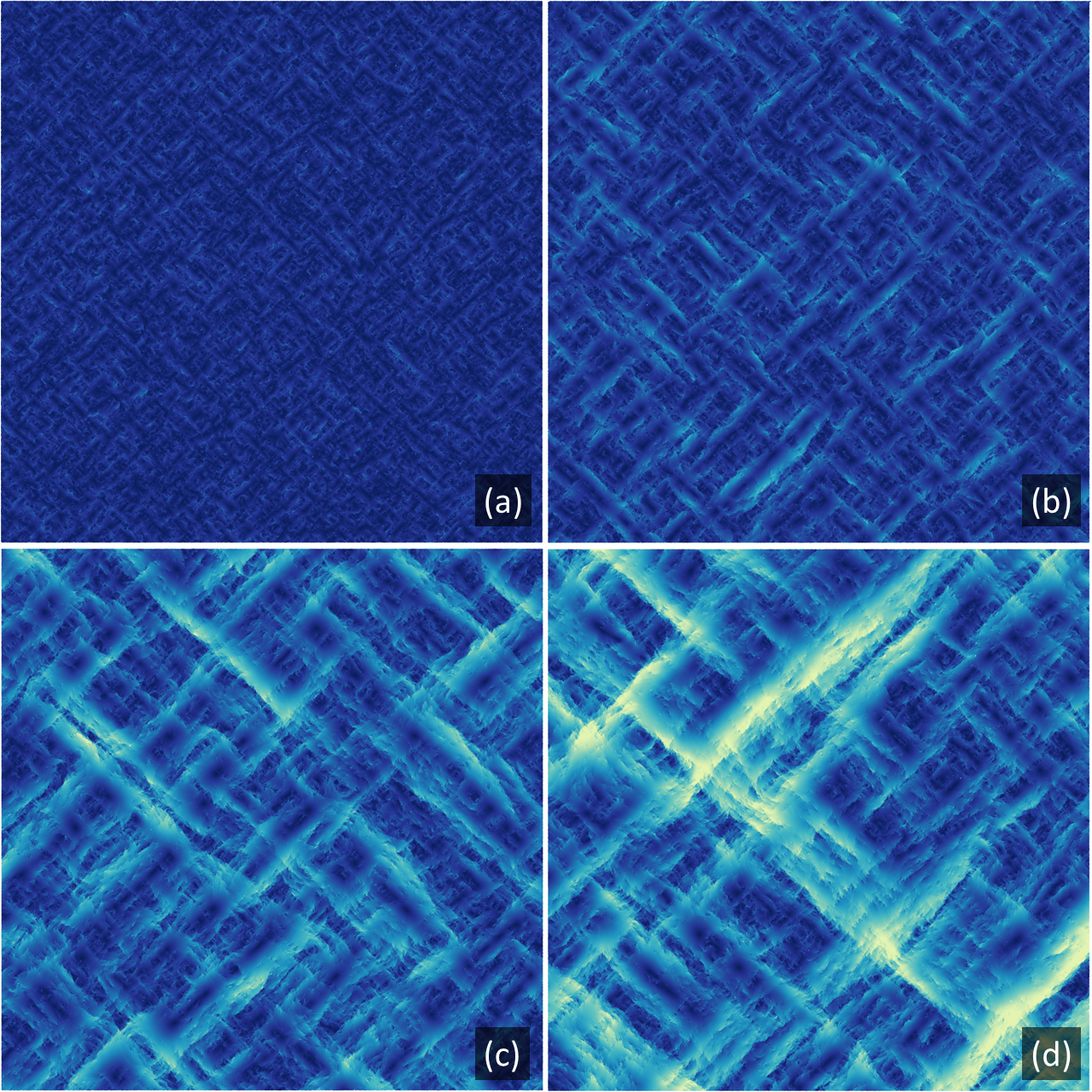}
	\caption{
	Rendered sections of 2D systems of size $L = 1753$ in steady-state flow at a strain rate of (a) $2\times10^{-4}$, (b) $2\times10^{-5}$, (c) $2\times10^{-6}$, and (d) $2\times10^{-7}$.
	Particles are colored by the magnitude of their accumulated nonaffine displacement over the previous interval of 2\% strain. 
	Dark blue corresponds to no displacement and bright, yellowish green corresponds to a displacement of 2.5. 
	This color scale cuts off high values of the displacement which can reach a maximum value of 2.9 and 3.6 in panels (c) and (d), respectively. 
	Systems are oriented with the compressive axis along the vertical direction and the extensional axis along the horizontal direction.
	In each panel, a square section of the periodic system is rendered with a side length of 1400. 
	Note that the square shape does not correspond to the KR box geometry and therefore a periodic copy of some particles may be rendered in some panels. 
	} \label{fig:NAD_Field}
	\end{center}
\end{figure*}

The changes in Fig. \ref{steady_flow} can be related to a
characteristic correlation length $\xi$ that diverges as $\sigma$ approaches the critical stress:
\begin{equation}
\xi \sim |\sigma - \sigma_c|^{-\nu} \ \ ,
\label{eq:xi}
\end{equation}
where $\nu$ is a critical exponent. Combining this expression with Eq. \eqref{eq:HB} yields:
\begin{equation}
\xi \sim \dot{\epsilon}^{-\nu /\beta} \ \ .
\label{eq:xi_epsilon}
\end{equation}
This length scale represents the maximum spatial range over which particles cooperatively rearrange during an avalanche.
In 2D, the divergence in $\xi$ is visible in the spatial correlations of the nonaffine displacement of particles as seen in Fig. \ref{fig:NAD_Field}.
During shear, avalanches displace particles along slip lines oriented along the direction of maximum shear stress (the $45^\circ$ diagonals between the compressive and extensional directions) as described in Refs. \cite{Maloney2008,Maloney2009}. 
As the rate decreases, displacements are correlated on longer length scales as larger avalanches are nucleated. 
This effect is discussed in further detail in Sec. \ref{sec:yield_diffusion}.

A finite system will be in the QS regime when $\xi > L$, so that avalanche size is limited only by the finite system dimensions.
The system will move to the finite strain rate (FSR) regime when the rate is large enough that $\xi < L$, and rate limits avalanche size.
At these high rates, the faster accumulation of stress reduces the time between avalanches leading to spatial and temporal overlap.
This overlap interferes with the evolution of avalanches and limits the maximum size.
Similar behavior has been identified in experimental studies of sheared granular packings \cite{Denisov2016}.
From Eq. \eqref{eq:xi_epsilon} the transition rate should scale as $\dot{\epsilon}_{QS}(L) \sim L^{-\beta/\nu}$. 
In Fig. \ref{fig:NAD_Field}(d), particle displacement is correlated on a length scale close to $L$ implying the system is strained at a rate close to $\dot{\epsilon}_{QS}(L)$.

Based on the scaling of avalanches in Sec. \ref{sec:qs_scaling}, we now provide bounds on the value of $\beta/\nu$ by considering the duration of avalanches.
Using Eq. \eqref{eq:RQS}, one can evaluate the fraction of the time that an avalanche is occurring in the QS regime, $f_\mathrm{act}$.
Conventionally, the duration of an avalanche $T_I$ scales as $T_I \sim E_I^{z/\alpha} \sim \ell_I^z$, where $z$ is the dynamic exponent and $\ell_I$ is the typical linear extent of an avalanche with energy $E_I$. 
Assuming that avalanches do not overlap in time, the total time for all avalanches in a unit strain is
\begin{equation}
	T_\mathrm{tot} = \sum_I T_I \sim \int dE R_{QS}(E,L) E^{z/\alpha} \ \ .
\end{equation}
The fraction of time where there are avalanches is then
\begin{equation}
	f_\mathrm{act}=\dot{\epsilon} T_\mathrm{tot}
	\sim  \dot{\epsilon} L^\gamma \int^{E_\mathrm{max}} dE E^{z/\alpha-\tau}
	\sim \dot{\epsilon} L^{y} \ \ ,
\label{eq:time}
\end{equation}
where $y\equiv \gamma+z +(1-\tau)\alpha$.
Using the scaling relation in Eq. \eqref{eq:scale},
\begin{equation}
y=d-\alpha+z \ \ .
	\label{eq:ydef}
\end{equation}

The QS regime should only be applicable when $f_\mathrm{act}$ is small.
Inserting $\dot{\epsilon}_{QS}(L) \sim L^{-\beta/\nu}$ into Eq. \eqref{eq:time} we see that $f_\mathrm{act}$ is only small at large $L$ if $\beta/\nu \geq y$.
The measured exponents described in the remainder of the paper are consistent with $y=\beta/\nu$, and thus with
a transition between QS and FSR regimes at a fixed value of $f_\mathrm{act}$.
If the inequality held, the fraction of activity would vanish as a power law as $L$ increased.

A separate upper bound for $\dot{\epsilon}_{QS}(L)$ can be obtained by determining
the rate where $f_\mathrm{act}$ approaches unity for a given $L$.
There is then constant activity and no quiescence in the system implying the system is well into the FSR regime.
Equation \eqref{eq:time} is only valid for small $f_\mathrm{act}$
because it ignores the possibility of temporal overlap between avalanches which becomes common as $f_\mathrm{act}$ rises to unity.

To determine whether the system has quiescent periods at a given rate, we evaluate the minimum and maximum kinetic energy, $K_\mathrm{min}$ and $K_\mathrm{max}$, during steady-state shear.
At low rates, the ratio $R_K \equiv K_\mathrm{min}/K_\mathrm{max}$ will be very small due to the contrast in $K$ during phases of activity versus inactivity as seen in Fig. \ref{steady_flow}(b). 
As $f_\mathrm{act}$ approaches unity, the kinetic energy has no time to decay between events and $R_K$ will rise and approach unity.
This transition is evident in Fig. \ref{steady_flow}(b), with no quiescent period for $\dot{\epsilon} \geq 2 \times 10^{-5}$.

\begin{figure}
\begin{center}
	\includegraphics[width=0.45\textwidth]{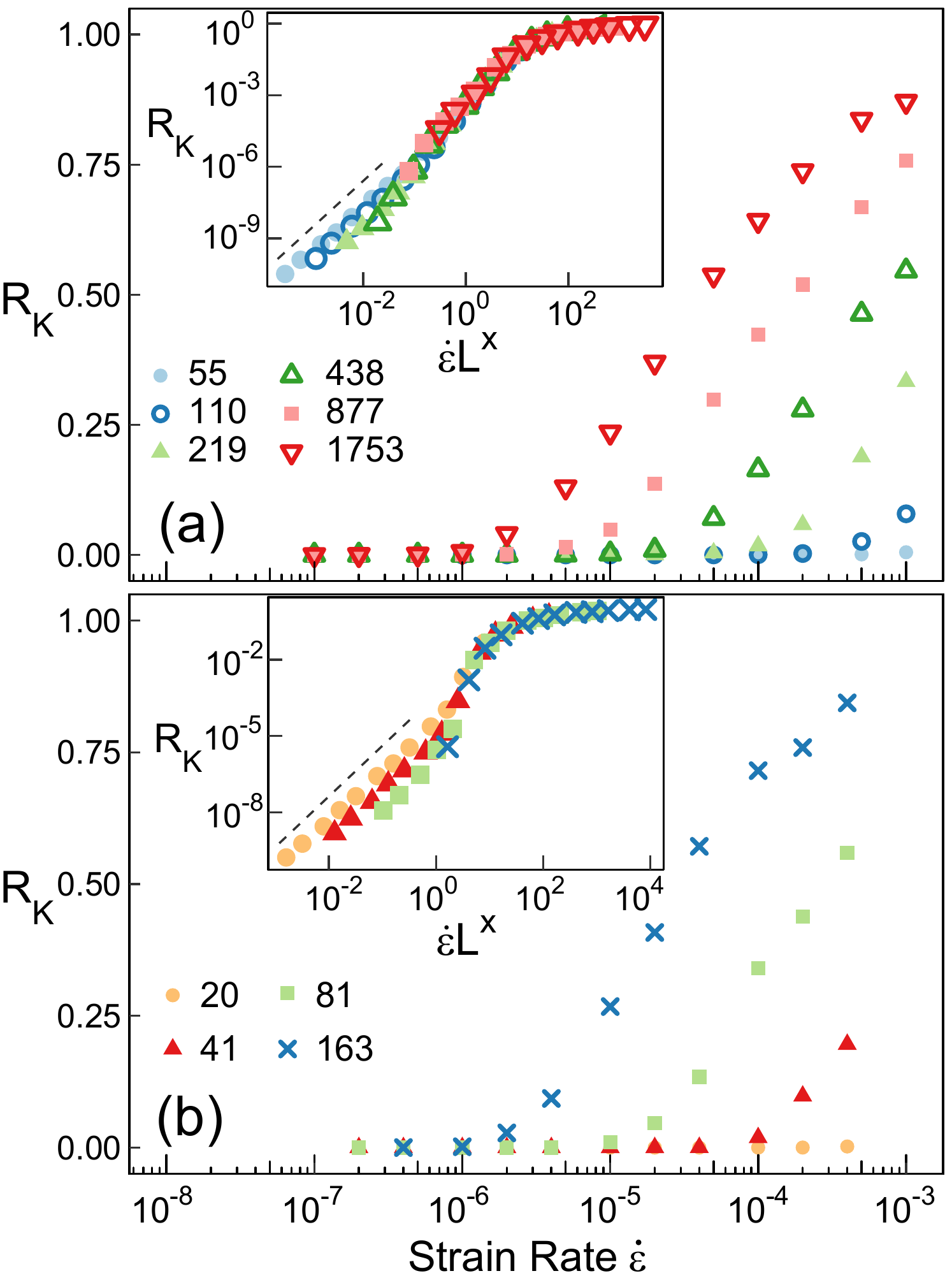}
	\caption{Ratio $R_K \equiv K_\mathrm{min}/K_\mathrm{max}$ plotted against $\dot{\epsilon}$ for (a) 2D and (b) 3D systems with the indicated $L$.
The insets show that scaling the rate by $L^x$ collapses data for different sizes with $x=d$. Dashed lines represent power laws with an exponent of 2.
	}
	\label{krat}
	\end{center}
\end{figure}

Figure \ref{krat} shows the variation of $R_K$ with rate for different $L$ in 2D and 3D.
As expected at low rates, $R_K \sim 0$. 
As $\dot{\epsilon}$ increases, $R_K$ rises rapidly and saturates at unity.
The rise occurs at lower rates as $L$ increases.
If this rate scales as $L^{-x}$, results for different $L$ should collapse when $R_K$ is plotted against $L^x \dot{\epsilon}$.
The insets in Fig. \ref{krat} show that this collapse is successful with $x=d$ at high and intermediate rates with some splay at the lowest rates.
At the lowest rates, $K_\mathrm{max}$ is nearly constant (Fig. \ref{steady_flow}(b)) and $R_K$ only reveals the variation of $K_\mathrm{min}$.
This is dominated by the background kinetic energy from nonaffine displacements giving $R_K \propto K_b \propto \dot{\epsilon}^2$.
Therefore in this limit, $R_K$ is not necessarily expected to collapse.
The additional variation of the prefactor with $L$ may provide information about the scaling of nonaffine displacements, but was not determined.

Uncertainties in fitting this exponent are of order 0.1 for this data.
Other measures of the onset of quiescence give the same scaling with smaller uncertainty as discussed in Appendix \ref{sec:yield_rmske}.
This appendix also provides a general argument for $x=d$ using other results from this paper.

The exponent $x$ describes the condition for $f_\mathrm{act}$ to be of order unity while $y$ is the condition for $f_\mathrm{act}$ to be small, implying $L^{-x} \geq L^{-y}$.
Since we find $x=d$ and $y=d+z-\alpha$,
\begin{equation}
d \leq d+z-\alpha \leq \beta/\nu \ \ .
	\label{eq:betabound}
\end{equation}
The first relation implies that $z \geq \alpha$.
The second relation provides a stricter bound on $\beta/\nu$. 
Studying noise spectra in the sibling paper, we find the opposite inequality, $y \ge \beta/\nu$, implying the presence of an equality \cite{Clemmer2021}.
This equality was also proposed using different arguments by Lin et al. \cite{Lin2014}.
In later sections we measure exponents and
show that $x<y$ but $y$ is consistent with $\beta/\nu$ in both 2D and 3D.
The values of $x$ and $y$ are very different in 2D,
implying that there must be substantial temporal overlap between avalanches in the FSR regime in order to allow for periods of quiescence.

%%%%%%%%%%%%%%%%%%%%%%%%%%%%%%%%%%%%%%%%%%%%%%%%%%%%%%%%%%%%
\section{Scaling of Steady-State Flow Stress}
\label{sec:steady}
%%%%%%%%%%%%%%%%%%%%%%%%%%%%%%%%%%%%%%%%%%%%%%%%%%%%%%%%%%%%

\begin{figure}
\begin{center}
	\includegraphics[width=0.45\textwidth]{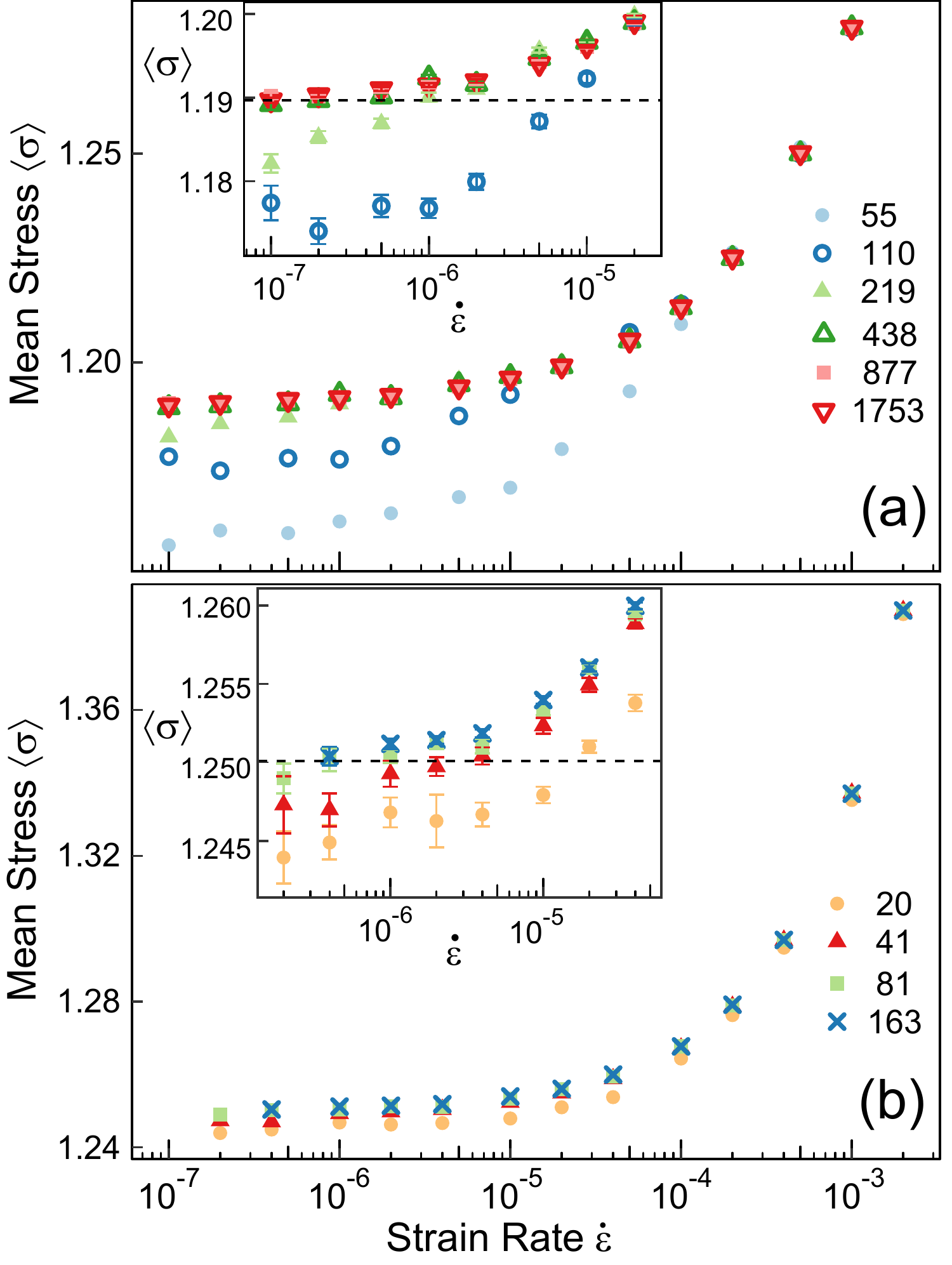}
	\caption{The average shear stress as a function of strain rate for systems of size $L$ indicated in the legends for (a) 2D and (b) 3D. The insets in each panel show a zoomed view of low rate data. The horizontal dashed lines indicate $\sigma_c =1.1897$ in 2D and $1.2501$ in 3D. 
	} 
	\label{average_stress}
	\end{center}
\end{figure}

The previous section described a transition between the QS and FSR regimes and obtained bounds on $\beta/\nu$. 
Here we look at the average stress as a function of strain rate and use results for different $L$ to determine $\beta$ and $\nu$. 
Figure \ref{average_stress} shows the variation of the average shear stress with rate for the indicated system sizes in 2D and 3D. Each point represents an average over ensembles as well as a strain interval in steady state.
Data is presented only up to $\dot{\epsilon} = 10^{-3}$ in 2D and $2 \times 10^{-3}$ in 3D because even these rates show deviations from critical scaling 
in some properties. 
In the sibling paper, no critical power-law is seen in temporal power spectra of the kinetic energy at these rates \cite{Clemmer2021}.

At high rates, the correlation length is small and results for different $L$ should converge.
For the 3D results in Fig. \ref{average_stress}(b), results for all $L$ lie on a common curve for $\dot{\epsilon}  \geq 10^{-3}$.
As the strain rate decreases, the $L=20$ results begin to fall below results for other sizes, indicating that the system is approaching the QS regime and $\xi > 20$.

The inset of Fig. \ref{average_stress}(b) shows how results for each $L$ deviate from those for larger $L$ as $\dot{\epsilon}$ decreases.
For each $L$ there is an approach to a limiting QS yield stress, $\sigma(\dot{\epsilon} = 0,L)$,
as $\dot{\epsilon} \rightarrow 0$.
The functional form of this rate dependence may include
non-critical behavior such as the scaling of $K_b$, the background kinetic energy.
As $L$ increases, $\sigma(0,L)$ increases towards the critical yield stress $\sigma_c$.
Similar behavior is seen for 2D systems in Fig. \ref{average_stress}(a).
The difference $\sigma(0,L) -\sigma_c$ is negative and should scale as $1/L^{1/\nu}$.
However, fits are complicated because fluctuations in stress, $\Delta \sigma$, are large
and statistical errors in the stress are of order 0.0005.
As discussed in the next section,  $\Delta \sigma \propto L^{-\phi}$ with $\phi < 1/\nu$,
making it difficult to resolve the critical region for large systems.

\begin{figure}
\begin{center}
	\includegraphics[width=0.45\textwidth]{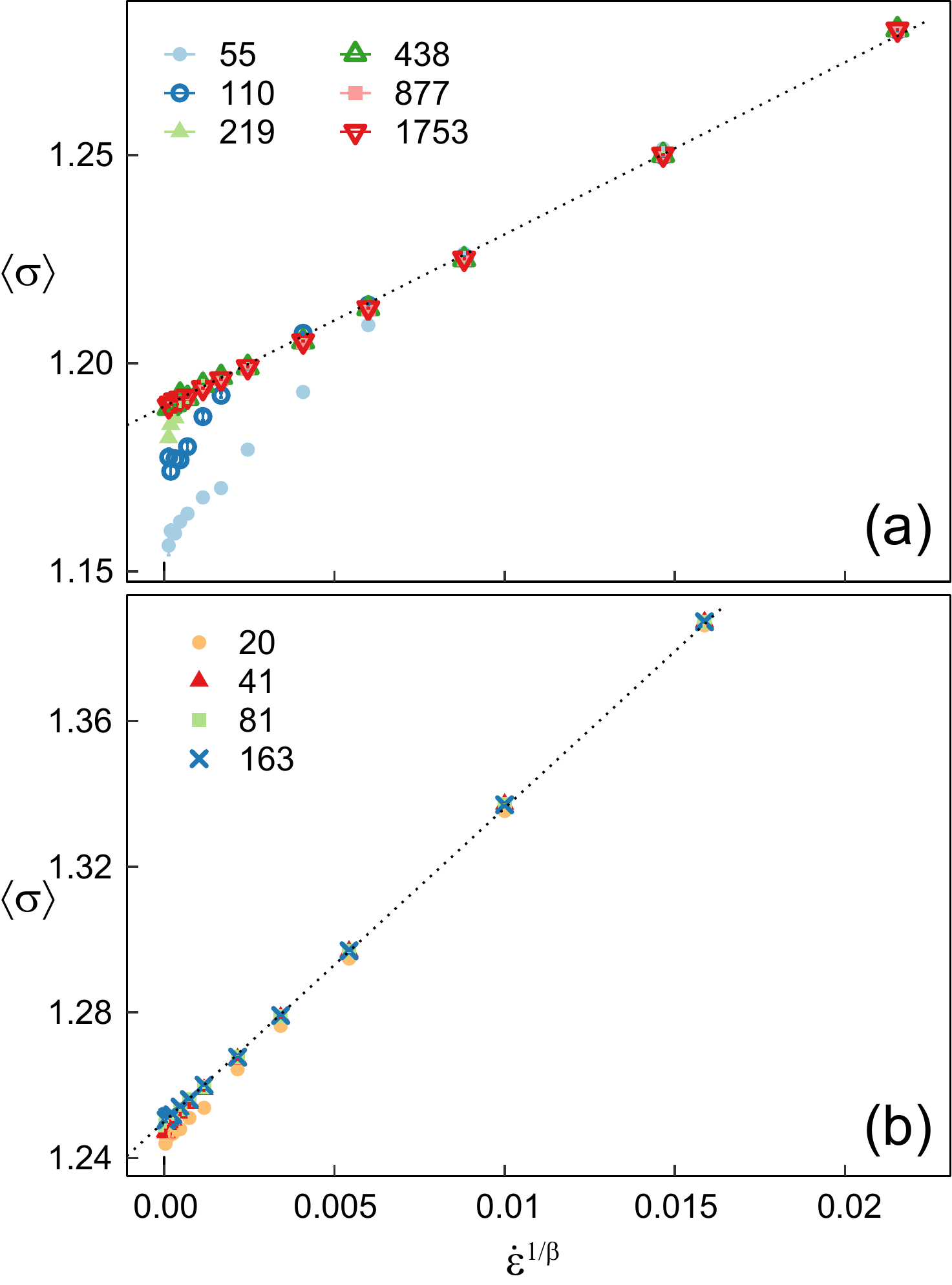}
	\caption{Stress plotted against strain rate to a power of $1/\beta$ chosen to give the linear scaling expected from Eq. \eqref{eq:HB}.
		(a) 2D data for the indicated $L$. A straight line (dashed) fit to data for $\dot{\epsilon}$ where $\xi < L$ gives $\beta=1.8$ and $\sigma_c=1.1897$.
(b) 3D data with a straight line fit giving $\beta=1.5$ and $\sigma_c=1.2501$.
} 
	\label{stress_beta}
	\end{center}
\end{figure}

The Herschel-Bulkley law in Eq. \eqref{eq:HB} applies to infinite systems.
Therefore, we first focus on data taken from systems at rates where $\xi < L$ so that finite-size effects are not important. 
For this subset of the data, $\langle \sigma \rangle$ does not depend on $L$ and thus is representative of an infinite system.
To reveal the power-law scaling, we plot $\sigma$ against $\dot{\epsilon}^{1/\beta}$ for the value of $\beta$ that produces the best straight line.
As shown in Fig. \ref{stress_beta}, the best fit gives $\beta=1.8 \pm 0.1$ and $\sigma_c=1.1897 \pm 0.0003$ in 2D and $\beta=1.50 \pm 0.05$ and $\sigma_c=1.2501 \pm 0.0003$ in 3D. The errorbars represent an estimate of the range of exponents that fit the region before finite-size effects set in.
These error bars are roughly estimated by accounting uncertainty in $\sigma_c$ and $\beta$ as well as uncertainty in what data is included in the fit. 
We note that both values of $\sigma_c$ are consistent with the average stress measured for our largest systems at the lowest rate as seen in the Fig. \ref{average_stress} insets.
The measured values of $\sigma_c$ and $\beta$ are used in subsequent plots and included in Table \ref{table:yield_exponents}.

The emergence of finite-size effects in Fig. \ref{average_stress} provides information about the rate dependence of $\xi$ that can be extracted using finite-size scaling techniques. As is typical in finite-size scaling theory, we assume that the only relevant length scales in the system are $L$ and $\xi$. Then the shear stress will depend only on the dimensionless scaling variable $L/\xi \propto L \dot{\epsilon}^{\nu/\beta}$ and $L$.
The resulting scaling ansatz can be written as
\begin{equation}
\langle \sigma \rangle -\sigma_c \sim L^{-1/\nu} g(\dot{\epsilon} L^{\beta/\nu}) \ \ ,
\label{eq:stress_scaling}
\end{equation}
where $g(x)$ is a universal scaling function.
For large $\dot{\epsilon}L^{\beta/\nu}$, finite-size effects are unimportant, and the critical scaling is recovered if $g(x) \sim x^{1/\beta}$ for $x \gg 1$.
For small $x$, $g$ must approach a constant that represents the shift of $\sigma(0,L)$ from $\sigma_c$.

\begin{figure}
\begin{center}
	\includegraphics[width=0.45\textwidth]{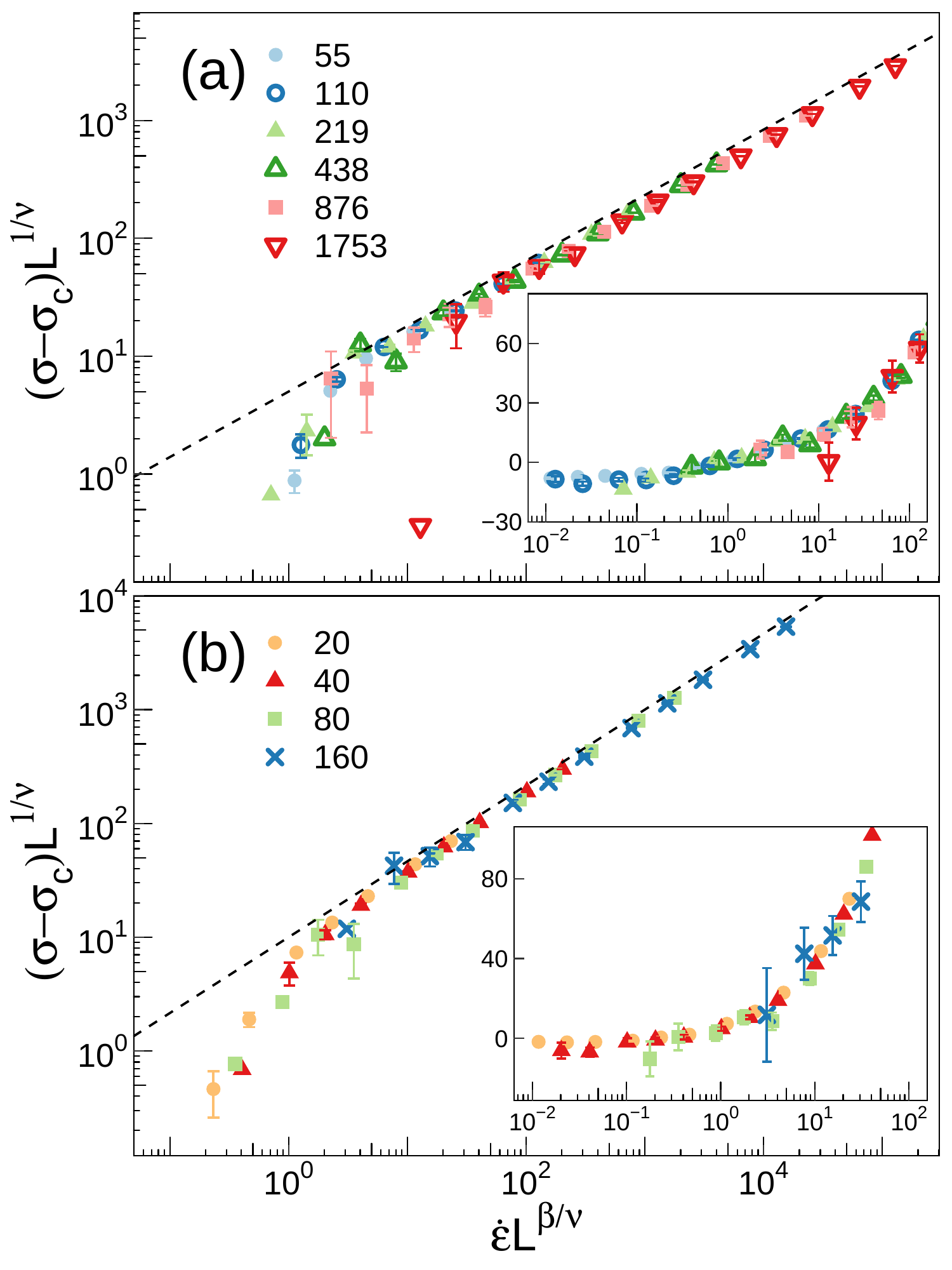}
	\caption{(a) The 2D data in Fig. \ref{average_stress}(a) is rescaled using the finite-size scaling relation in Eq. \eqref{eq:stress_scaling} with values of $\beta = 1.8$, $\nu = 0.72$, and $\sigma_c = 1.1897$. The dashed line represents a power-law with an exponent $1/\beta = 0.56$.
(b) 3D data from Fig. \ref{average_stress}(b) rescaled with $\beta = 1.5$, $\nu = 0.48$, and $\sigma_c = 1.2501$. The dashed line is a power-law with exponent $1/\beta = 0.67$. The insets in both panels include a zoomed in view of the same scaled data on a linear-log scale to show the collapse for $\langle \sigma \rangle < \sigma_c$.
} 
	\label{stress_scaled}
	\end{center}
\end{figure}

Equation \eqref{eq:stress_scaling} implies that results for all $L$ should
collapse if $(\langle \sigma \rangle -\sigma_c) L^{1/\nu}$ is plotted against $\dot{\epsilon}L^{\beta/\nu}$.
Figure \ref{stress_scaled} shows collapses for both 2D and 3D data.
As noted above, the statistical uncertainties in $\sigma$ are of order 0.0005 for low rates because of the large run times required to get better statistics. 
Errors are indicated when they are larger than the symbol size.
It is also unclear what range of rates remains in the critical regime.
Using the values of $\sigma_c$ and $\beta$ determined from Fig. \ref{stress_beta}, we found $\nu=0.72 \pm 0.06$ in 2D and $0.48 \pm 0.06$ in 3D.
This corresponds to $\beta/\nu = 2.5 \pm 0.3$ in 2D and $3.1\pm 0.4$ in 3D,
which are consistent with $\beta/\nu >d$.
This estimate of $\beta/\nu$ is consistent with the scaling of other system properties with strain rate such as the effective particle diffusion in 2D discussed in Sec. \ref{sec:yield_diffusion} and the noise spectra discussed in the sibling paper \cite{Clemmer2021}.

Figure \ref{stress_scaled} implies that the transition between QS and FSR regimes happens at $\dot{\epsilon} \sim L^{-\beta/\nu}$.
In the previous section we argued that the QS regime had to end by a rate
$\dot{\epsilon} \sim L^{-y}$, implying $y=d+z-\alpha \leq \beta/\nu$.
The numerical results summarized in Table \ref{table:yield_exponents} are consistent with the equality for both 2D and 3D.
This relation is motivated by other arguments below.

%%%%%%%%%%%%%%%%%%%%%%%%%%%%%%%%%%%%%%%%%%%%%%%%%%%%%%%%%%%%
\section{Fluctuations in the Shear Stress}
\label{sec:yield_deviate}
%%%%%%%%%%%%%%%%%%%%%%%%%%%%%%%%%%%%%%%%%%%%%%%%%%%%%%%%%%%%

Further information about the critical exponents can be obtained by considering fluctuations in the system.
Reference \cite{Salerno2013} examined the scaling of the standard deviation
of the stress,
$\Delta \sigma \equiv \sqrt{\langle \sigma^2 \rangle- \langle \sigma \rangle^2}$,
in the QS regime.
The results are consistent with
\begin{equation} 
\Delta \sigma_\mathrm{QS} \sim L^{-\phi} \ \ ,
\label{eq:deltas_qs}
\end{equation}
where $\phi$ is another critical exponent and
the subscript QS indicates the relation holds in the quasistatic
regime where $\xi > L$.
The value of $\phi$ reflects the strength of correlations in the stress of the system and one can define two upper bounds for a $d$ dimensional system \cite{Salerno2013}.
If there are no spatial correlations in the stress field and stress-drops are associated with a finite cluster of particles, incoherent addition would imply $\Delta \sigma \sim L^{-d/2}$.
If there are correlations in the stress in different regions, it could slow the decrease in fluctuations with $L$.
This implies $\phi \leq d/2$.
The scale of fluctuations must also be at least as large as the
magnitude of the stress drop during the largest avalanche. 
The energy released in the largest avalanche scales as $E_\mathrm{max} \sim L^\alpha$ implying 
a change in the intensive stress of $\Delta \sigma \sim L^{(\alpha-d)}$.
Since this is a lower bound on fluctuations, $\phi \le d-\alpha$.

\begin{figure}
\begin{center}
	\includegraphics[width=0.45\textwidth]{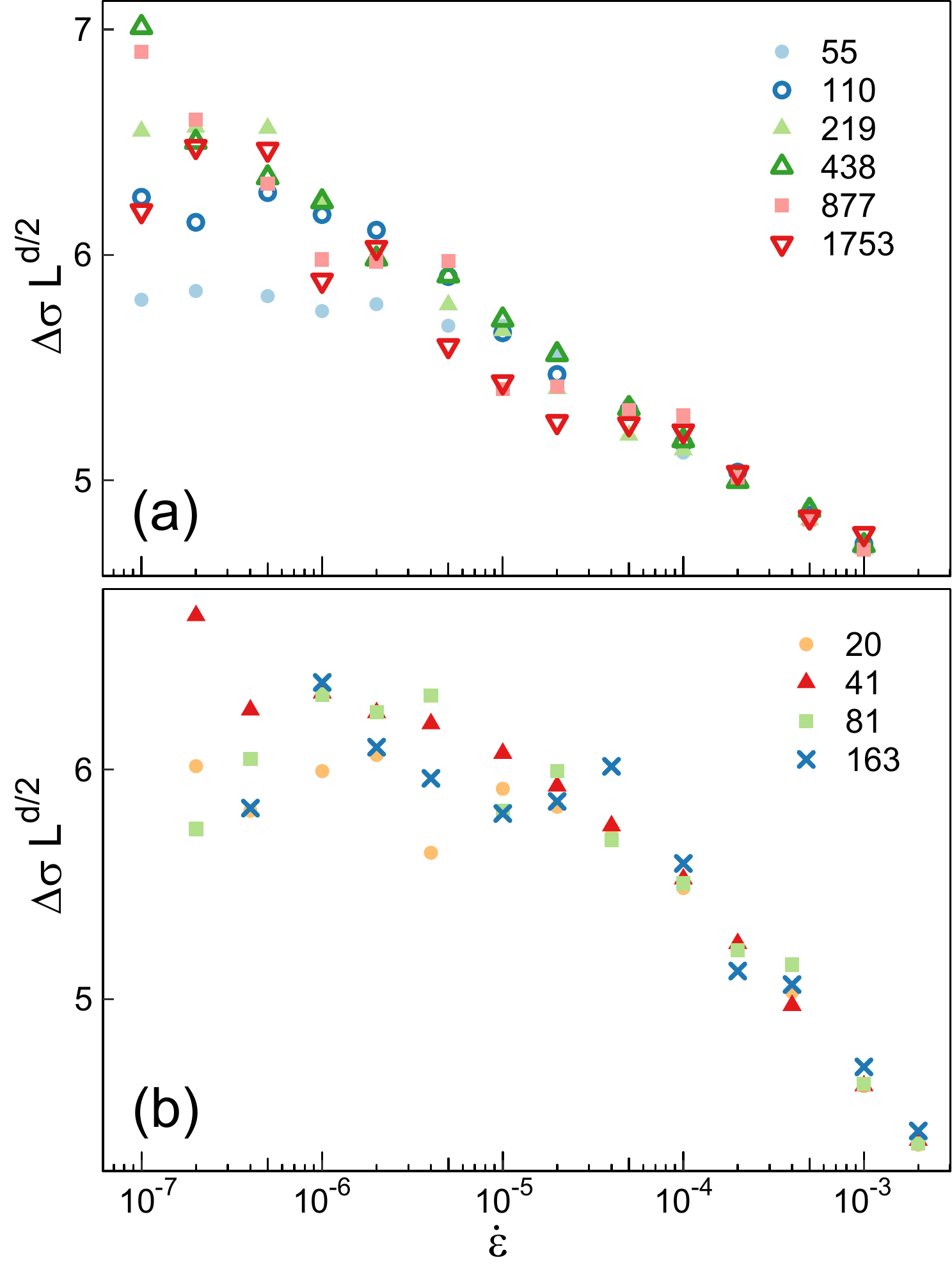}
	\caption{ Standard deviation of stress multiplied by $L^{d/2}$ as a function of rate for the indicated $L$ in (a) 2D and (b) 3D.
	In the QS regime, statistical error bars are of order 0.1 to 0.2. At high rates, averaging over independent regions causes errorbars to decrease to of order the symbol size.}
	\label{variance}
	\end{center}
\end{figure}

At finite strain rates, regions of size $\xi^d$ are uncorrelated and their contributions to $\sigma$ will add incoherently.
Near the critical point, the fluctuations in stress within each subregion should
scale as $\Delta \sigma_\xi \sim \xi^{-\phi}$. The number of these uncorrelated subregions will scale as $N_\xi \sim (L/\xi)^d$. 
Therefore, fluctuations in the total stress scale as $\Delta \sigma \sim \Delta \sigma_\xi N_\xi^{-1/2} \sim L^{-d/2} \xi^{d/2-\phi}$. This expression can be reexpressed in terms of strain rate using Eq. \eqref{eq:xi_epsilon}:
\begin{equation}
\Delta \sigma_\mathrm{FSR} \sim L^{-d/2} \dot{\epsilon}^{(\phi-d/2)\nu/\beta} \ \ ,
\label{eq:deltas_fsr}
\end{equation}
where the FSR subscript emphasizes that this scaling holds in the finite strain rate regime.
Note that in the special case of $\phi=d/2$ the QS and FSR regimes both scale as $L^{-d/2}$ and fluctuations
are independent of rate.

In $d = 3$, Salerno and Robbins measured $\alpha = 1.1 \pm 0.1$ in the overdamped limit implying $\phi$ is more strictly bounded by $d/2$ \cite{Salerno2013}.
This upper limit was found to be consistent with their actual measurement of $\phi = 1.5 \pm 0.2$.
From the above equations, this implies that $\Delta \sigma L^{3/2}$ should be nearly independent of system size and rate.
Fig. \ref{variance}(b) confirms this prediction.
In the QS regime, $\Delta \sigma L^{3/2}$ is near 6 for all $L$. 
Testing other scaling exponents indicates that
the QS results are consistent with $\phi=1.47 \pm 0.07$ in agreement with Ref. \cite{Salerno2013}, but with tighter error bars.
In the FSR regime,
all of the results collapse within statistical errors.
While there is a small decrease in $\Delta \sigma L^{3/2}$ with increasing rate
that might suggest $\phi < d/2$, the change is only 30\% over
more than two decades in rate.
This also implies that $\phi=d/2$ with an uncertainty of less than 0.1.

For $d = 2$, past results gave $\alpha = 0.9 \pm 0.05$ in the overdamped limit \cite{Salerno2013}.
Once again, this gives a larger upper bound than $d/2$, implying that $\phi=d/2=1$ 
which agreed with the measured value of $\phi = 1.0 \pm 0.1$ \cite{Salerno2013}.
Figure \ref{variance}(a) shows a plot of $\Delta \sigma L$ against rate for multiple system sizes.
Results in the QS regime collapse for large $L$, confirming that $\phi=d/2$.
There is a small drop in $\Delta \sigma L$ as $L$ decreases to 110 and 55 that
is consistent with deviations from critical scaling in small systems.
All of the results collapse in the FSR regime.
As in 3D, there is a small decrease with increasing rate,
but the results are consistent with $\phi=d/2=1$ with an uncertainty of less than 0.1.

\begin{figure}
\begin{center}
	\includegraphics[width=0.45\textwidth]{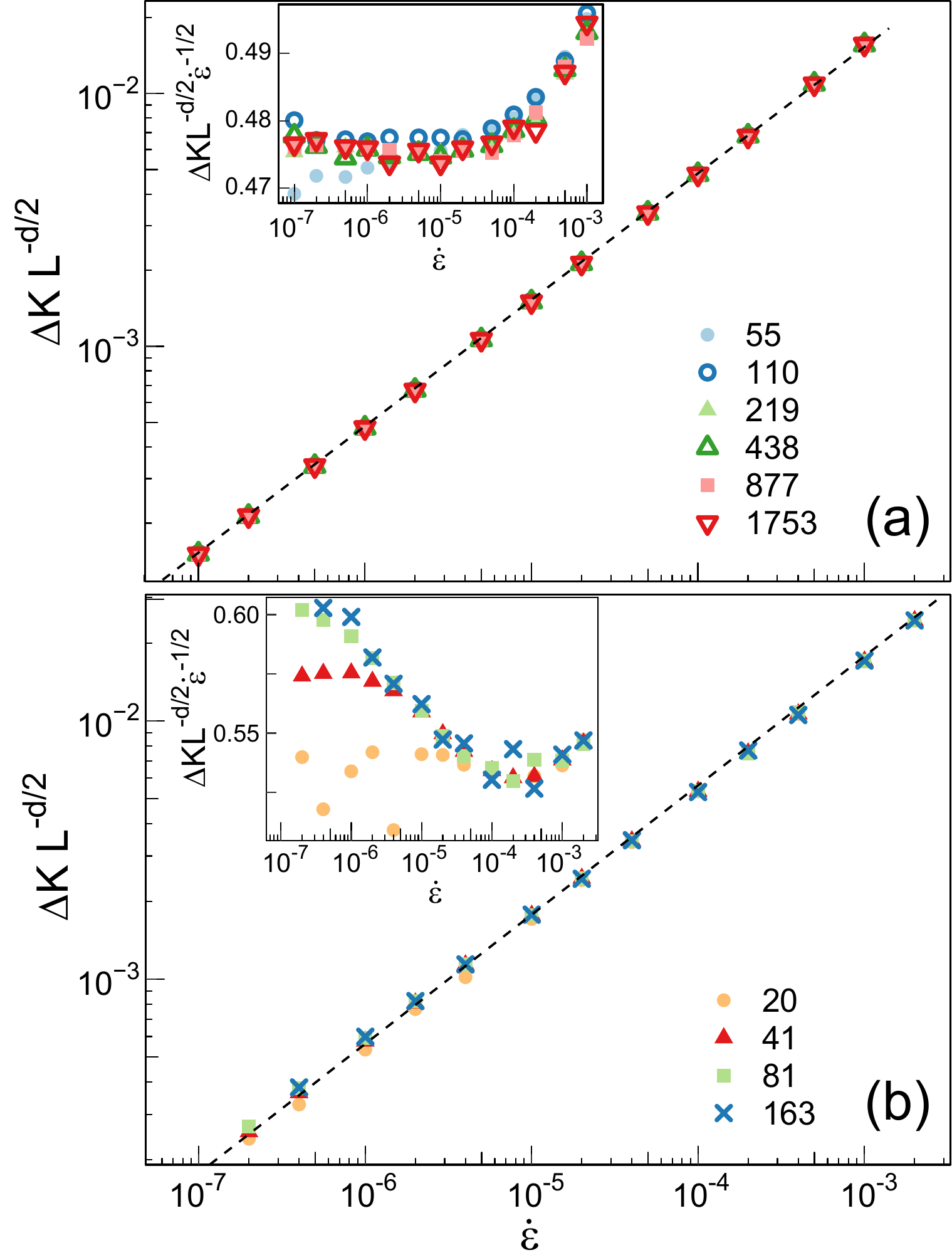}
	\caption{ Standard deviation of total kinetic energy multiplied by  $L^{-d/2}$ as a function of rate for the indicated $L$ in (a) 2D and (b) 3D. 	Dashed lines in both panels show power laws with exponents of 1/2. Insets in each panel show the same data divided by $\dot{\epsilon}^{1/2}$
	to test for any deviations from scaling.
} 
	\label{kevariance}
	\end{center}
\end{figure}

As noted in Sec. \ref{sec:qs_scaling}, there is an approximate correspondence between $K/L^d$ and $d \sigma/dt$. 
It is therefore reasonable to expect that $\Delta K/L^d$ will scale with the same power of $L$ as $\Delta \sigma $ in Eqs. \eqref{eq:deltas_qs} and \eqref{eq:deltas_fsr}. 
Figure \ref{kevariance} confirms that multiplying $\Delta K/L^d$ by $L^{d/2}$ collapses data for different $L$ in both 2D and 3D.
As illustrated by the dashed lines, for all systems
$\Delta K /L^{d/2}$ increases approximately as $\dot{\epsilon}^{1/2}$. 
The insets in the figure show $\Delta K/L^{d/2} \dot{\epsilon}^{-1/2}$ varies only a few percent over the entire range of rates.
At quasistatic rates, a minor deviation ($< 10$ \%) in scaling with $L$
is identifiable in 3D. 
This could either be a correction to scaling or indicate a decrease in $\phi$ from $1.5$ by less than 0.06.

In the QS regime, the dependence on strain rate can be easily explained.
As the rate continues to decrease, the same sequence of avalanches evolves in a given strain increment.
Thus the integral of $K(t)$ or $K^2(t)$ over that strain will be constant.
However the mean value will be normalized by the total time to achieve the strain and thus any moment, such as $\langle K \rangle$ and $\langle K^2 \rangle$, will be proportional to $\dot{\epsilon}$.
The variance $\Delta K^2 \equiv \langle K^2 \rangle - \langle K \rangle ^2$ will be dominated by $\langle K^2 \rangle \propto \dot{\epsilon}$ 
since the second term scales as $\dot{\epsilon}^2$.
The scaling is worked out in more detail in Appendix \ref{sec:yield_rmske} and gives 
\begin{equation}
\Delta K_\mathrm{QS} \sim L^{d/2} \dot{\epsilon}^{1/2}
\label{eq:var_ke}
\end{equation}
where the QS subscript emphasizes that this relation holds only in the quasistatic regime. 
In the FSR regime, one could imagine that similarly $\Delta K_\mathrm{FSR}/L^d \sim \Delta \sigma_\mathrm{FSR} \dot{\epsilon}^{1/2}$ as suggested by the data in Fig. \ref{kevariance}.

%%%%%%%%%%%%%%%%%%%%%%%%%%%%%%%%%%%%%%%%%%%%%%%%%%%%%%%%%%%%
\section{Particle Diffusion}
\label{sec:yield_diffusion}
%%%%%%%%%%%%%%%%%%%%%%%%%%%%%%%%%%%%%%%%%%%%%%%%%%%%%%%%%%%%

As a system is strained, particles plastically rearrange and exchange neighbors during avalanches. 
A particle's accumulated motion due to plasticity is measured by the nonaffine displacement, $\Delta \vec{r}$.  
This specifically does not include the distance a particle has moved due to the affine motion from the box shear. 
Previous studies have identified that the mean-squared nonaffine displacement, $\langle \Delta  \vec{r}^2 \rangle$, grows linearly with strain in 2D \cite{Lemaitre2007, Maloney2008} and 3D \cite{SalernoThesis}. 
One can therefore define an effective diffusion coefficient $D$, the rate of increase in $\langle \Delta  \vec{r}^2 \rangle$ per unit strain, to quantify the rate of particle transport. 
In this section, we study the critical scaling of $D$, test our previously measured value of $\beta/\nu$, and identify novel geometric effects.

In the QS regime, it has been observed that $D$ grows linearly with $L$ in 2D \cite{Lemaitre2007, Maloney2008}. 
Two explanations have been proposed for this observation. 
The first is that the maximum span of a slip line, $L$, determines the rate of particle diffusion, assuming $\alpha = 1$ \cite{Lemaitre2007}. 
The second is based on the observation that plastic deformation is correlated over an interval of strain that scales as $L^{-1}$ \cite{Maloney2008}.  

In contrast, an alternate scaling theory was proposed by Tyukodi, Vandembroucq, and Maloney \cite{Tyukodi2018, Tyukodi2019} who recognized that slip lines are created by fractal avalanches.
The authors estimated the rate of slip line formation by approximating the strain released by a system spanning slip line.
Combining this rate with the expected displacement produced by a slip line, they argued the diffusion grows as $L^{2-\alpha}$ for their 2D EPM systems.
In the QS regime, Tyukodi et al. found the diffusion grew as $L^{1.05}$ implying that $\alpha = 0.95$.

Based on their arguments, we propose a similar scaling relation which accounts for a power-law distribution of avalanche sizes. 
This allows us to extend the theory to the FSR regime.
First we consider the displacement field created by a slip line.
Particle displacements decay approximately linearly with the distance from the slip line \cite{Maloney2008}.
As argued in Refs. \cite{Maloney2008, Tyukodi2018, Tyukodi2019}, this implies each slip line $I$ will contribute a factor of $\sim \delta^2/12$ to the accumulated mean-square nonaffine displacement, $\langle \Delta \vec{r}^2 \rangle$. Here $\delta$ is the typical change in displacement across the slip line. 
Notably, this does not depend on the length of the slip line. 

One could imagine every avalanche of magnitude $E \sim \ell^\alpha$ creates a local slip line of length $\ell$ which displaces particles within a local region of size $\ell^d$. 
This will increase the local mean-squared nonaffine displacement by a factor of 
$\delta^2/12$ and globally increase $\langle \Delta \vec{r}^2 \rangle$ by a factor of $(\ell/L)^d \delta^2/12$.
The diffusion rate can then be estimated as:
\begin{equation}
D = \frac{\langle \Delta \vec{r}^2 \rangle}{\Delta \epsilon} \sim \frac{1}{\Delta \epsilon} \frac{\delta^2}{12} \sum_I \left(\frac{\ell_I}{L}\right)^d 
\end{equation}
where the summation is over all avalanches in the strain interval $\Delta \epsilon$.
This sum can be rewritten as
\begin{equation}
D \sim \frac{\delta^2}{12 L^{d}} \int_0^{E_\mathrm{max}} dE E^{d/\alpha} R_{QS}(E,L) 
\end{equation}
using the nucleation rate of avalanches from Sec. \ref{sec:qs_scaling}.
Dropping constant prefactors and applying Eq. \eqref{eq:RQS}, we find
\begin{align}
\begin{split}
D 
  &\sim L^{\gamma-d} \int_0^{E_\mathrm{max}} dE E^{d/\alpha} E^{-\tau}\\
  & \sim L^{\gamma + \alpha(1-\tau)} \\
  & \sim L^{d-\alpha}
\label{eq:diffusion_scaling}
\end{split}
\end{align}
where the final relation uses the scaling relation for $\gamma$ in Eq. \eqref{eq:scale}.
This expression is similar to the scaling relation derived by Tyukodi et al. \cite{Tyukodi2018, Tyukodi2019}.

It is important to note that the integral in Eq. \eqref{eq:diffusion_scaling} is dominated by system-spanning avalanches. 
Therefore, the same result is reached if one only considers slip lines of length $L$.
Each slip line would contribute a constant factor of $\delta^2/12$ to $\langle \Delta \vec{r}^2 \rangle$ so the diffusion would simply be proportional to the nucleation rate of system-spanning avalanches or $L^{d-\alpha}$, as derived in Ref. \cite{Clemmer2021}.

This scaling theory is easily extended to the FSR regime. 
In this limit, the distribution of avalanches is cut off at $E_\mathrm{max}^{FSR} \sim \xi^\alpha$. 
Space is partitioned into $(L/\xi)^d$ independent regions each of which contributes to the global diffusion by a factor of:
\begin{equation}
 \frac{\delta^2 \xi^\gamma}{12 L^{d}} \int_0^{E_\mathrm{max}^{FSR}} dE E^{d/\alpha} E^{-\tau} \sim \frac{\xi^{2d -\alpha}}{L^d}\\
\end{equation}
similar to Eq. \eqref{eq:diffusion_scaling}.
Summing all contributions, the total diffusion of the system will scale as:
\begin{equation}
D \sim \left(\frac{L}{\xi}\right)^d \frac{\xi^{2d -\alpha}}{L^d} \sim \xi^{d-\alpha} \sim \dot{\epsilon}^{(\alpha-d)\nu/\beta}
\label{eq:diffusion_fsr}
\end{equation}
where the final relation follows from Eq. \eqref{eq:xi_epsilon}.

We now test these relations using results from simulations. 
We focus first on the diffusion in 2D then discuss 3D at the end of this section.
Simulations were split into intervals of 5\% strain and the cumulative nonaffine displacement was averaged over increments of 0.1\% strain for each interval. 
A least mean squares linear regression was then used to fit the data for each interval to calculate a diffusion coefficient at that particular value of the strain.
We also tested using instantaneous values of the nonaffine displacement at each 0.1\% strain increment instead of averaged quantities, but found no significant difference.

We first consider the effect of the global system geometry on the diffusion. 
In KR boundary conditions, the simulation box is regularly remapped with a period of strain approximately equal to $\epsilon_{KR} \approx 0.96$. 
This remapping allows deformation to reach arbitrarily large elongational strains. 
We therefore define $\epsilon_M = \epsilon~\mathrm{mod}~\epsilon_{KR}$ as the current location of the simulation box in strain space. 
Different values of $\epsilon_M$ correspond to different lattice vectors of the simulation box.

\begin{figure}
\begin{center}
	\includegraphics[width=0.45\textwidth]{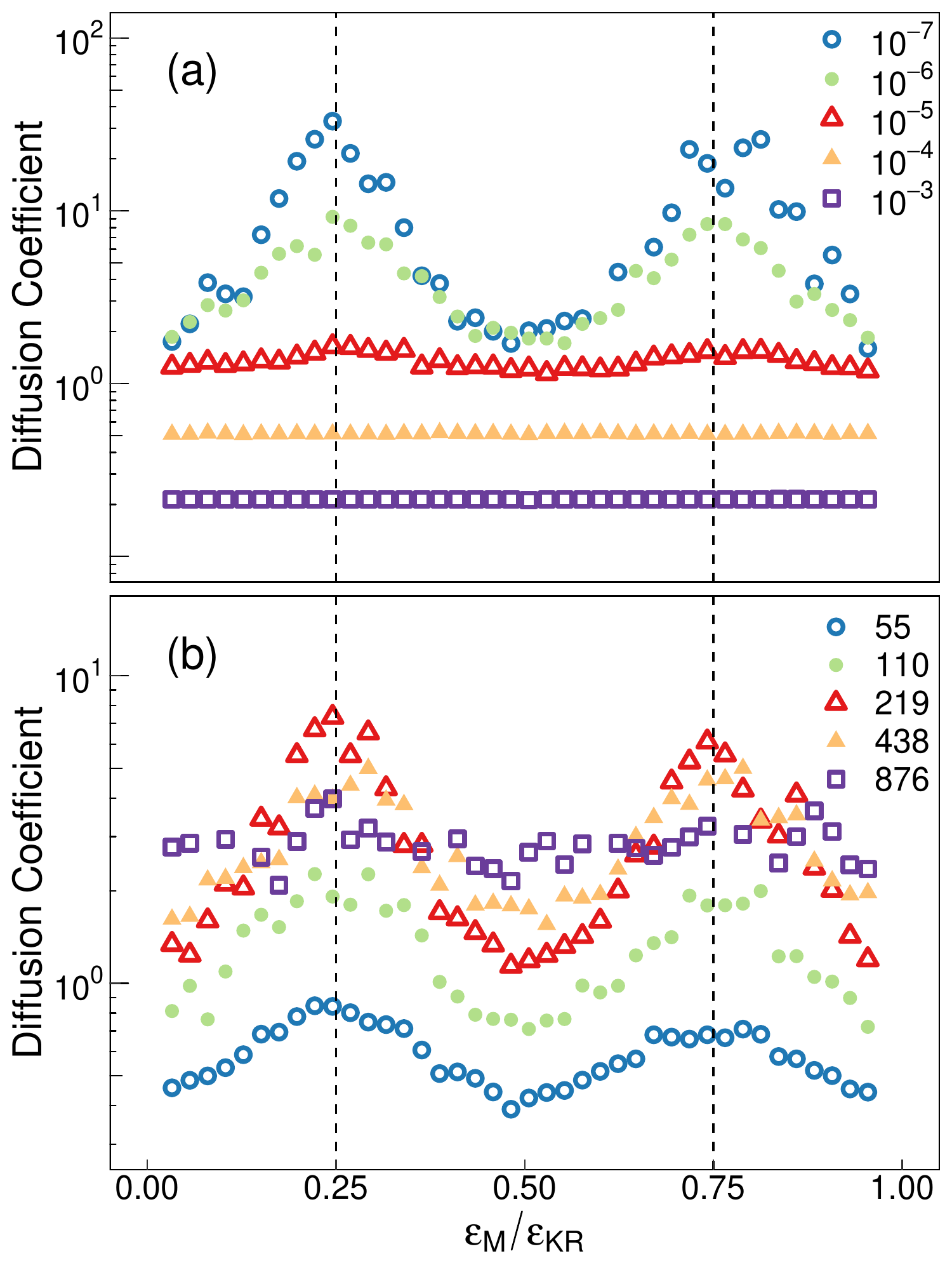}
	\caption{
	(a) The average diffusion coefficient plotted as a function of $\epsilon_{M}/\epsilon_{KR}$ for 2D systems of $L = 438$ strained at the indicated rates. 
(b) Similar data is plotted at a fixed rate of $2\times 10^{-6}$ for the indicated system sizes.
Vertical dashed lines in both panels represent values of $\epsilon_M = 1/4 \epsilon_{KR}$ and $3/4 \epsilon_{KR}$. 
	} 
	\label{fig:diffusion_strain}
	\end{center}
\end{figure}

In Fig. \ref{fig:diffusion_strain}(a), the average diffusion coefficient is plotted as a function of $\epsilon_M/\epsilon_{KR}$ for a system of size $L = 438$ for the indicated strain rates. 
At a high rate of $\dot{\epsilon} \ge 10^{-4}$, systems are in the FSR regime and and there is no dependence on $\epsilon_M$. 
As the strain rate is decreased, diffusion increases. 
At a rate of $10^{-5}$, small undulations in the diffusion rate emerge and continue to grow as the strain rate continues to drop. 
At rates of $\dot{\epsilon} \le 10^{-6}$, the diffusion coefficient clearly peaks at values of $\epsilon_M = 1/4 \epsilon_{KR}$ and $3/4 \epsilon_{KR}$. 
In Fig. \ref{fig:diffusion_strain}(b), data is shown for different system sizes at a fixed rate of $2\times 10^{-6}$.  
Fluctuations are only seen in systems with $L < 876$ implying they are a finite-size effect. 
The amplitude of fluctuations also increases with decreasing rate and increasing system size.
Note that no other system properties discussed in this article depended on $\epsilon_{M}$.

These fluctuations are a geometric effect that corresponds to the periodic boundary conditions.
Shear stress is maximized along lines oriented $45^\circ$ between the compressive and extensional axes.
Avalanches therefore preferentially orient along these directions as seen in Fig. \ref{fig:NAD_Field}.
Near the QS regime, the linear span of an avalanche approaches the size of the system. 
If an avalanche crosses a periodic boundary, it will wrap back around and emerge on the other side of the simulation cell.
Strains of $\epsilon_M = 1/4 \epsilon_{KR}$ and $3/4 \epsilon_{KR}$ correspond to special box geometries where such slip lines will perfectly wrap around the box, enhancing particle diffusion.

\begin{figure}
\begin{center}
	\includegraphics[width=0.42\textwidth]{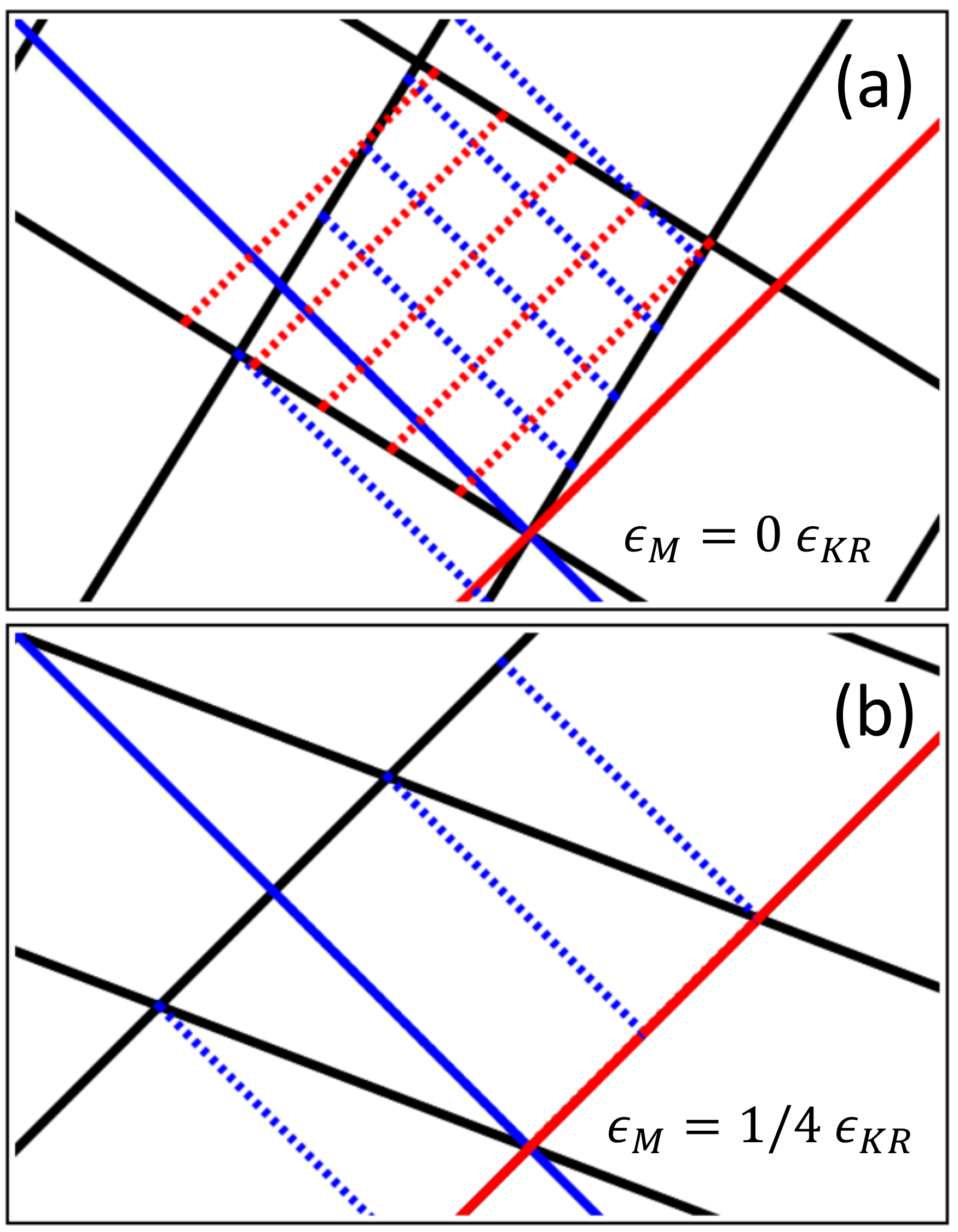}
	\caption{
	Periodic unit cells (black) using the KR boundary conditions at a value of (a) $\epsilon_M = 0 \epsilon_{KR}$ and (b) $1/4 \epsilon_{KR}$.  
	The lattice has been rotated into the laboratory frame such that the principal stress vectors are aligned with the vertical (compressive) and horizontal (extensional) dimensions of the panel. 
	Solid blue and red lines correspond to the directions of maximal shear stress ($45^\circ$ diagonals). 
	Dashed blue and red lines correspond to periodic remappings of these lines. 
	Note that the dashed lines in (a) do not remap onto an existing line while dashed lines in (b) do remap onto existing lines.
	There are no red dashed lines in (b) as the red line is parallel to one set of the black lattice vectors and therefore perfectly remaps.
	} 
	\label{fig:kr_grid}
	\end{center}
\end{figure}

For instance, at a strain of $\epsilon_M = 0 \epsilon_{KR}$, $45^\circ$ slip lines do not wrap perfectly around the box as seen in Fig. \ref{fig:kr_grid}(a).
In fact, spanning avalanches may suppress diffusion.
On either side of a slip line, particles will flow in opposite directions. 
Therefore, slip lines that do not coincide with themselves after wrapping will force anti-parallel flow between the original slip line and its periodic image.
Interestingly in Fig. \ref{fig:diffusion_strain}(b), diffusion for $L = 438$ is suppressed at a strain of $\epsilon_M = 0 \epsilon_{KR}$ and $1/2 \epsilon_{KR}$ relative to the average diffusion of $L = 876$ which is independent of $\epsilon_M$ as it is in the FSR regime.
This suggests there may be maximal deconstructive interference at these geometries.

In contrast at strains of $\epsilon_M = 1/4 \epsilon_{KR}$ and $3/4 \epsilon_{KR}$, large slip lines periodically remap onto themselves.
This is illustrated for  $\epsilon_M = 1/4 \epsilon_{KR}$ in Fig. \ref{fig:kr_grid}(b) as one of the $45^\circ$ diagonals is parallel to one of the lattice vectors of the simulation cell and the other diagonal remaps onto itself after crossing one periodic image (maximizing the distance between remappings of a slip line).
Similar behavior is seen at $\epsilon_M = 3/4 \epsilon_{KR}$ (not shown).
At low strain rates, there is pronounced accumulation of nonaffine displacement along these directions at these special strains which would be visible in images like Fig. \ref{fig:NAD_Field}.

\begin{figure}
\begin{center}
	\includegraphics[width=0.45\textwidth]{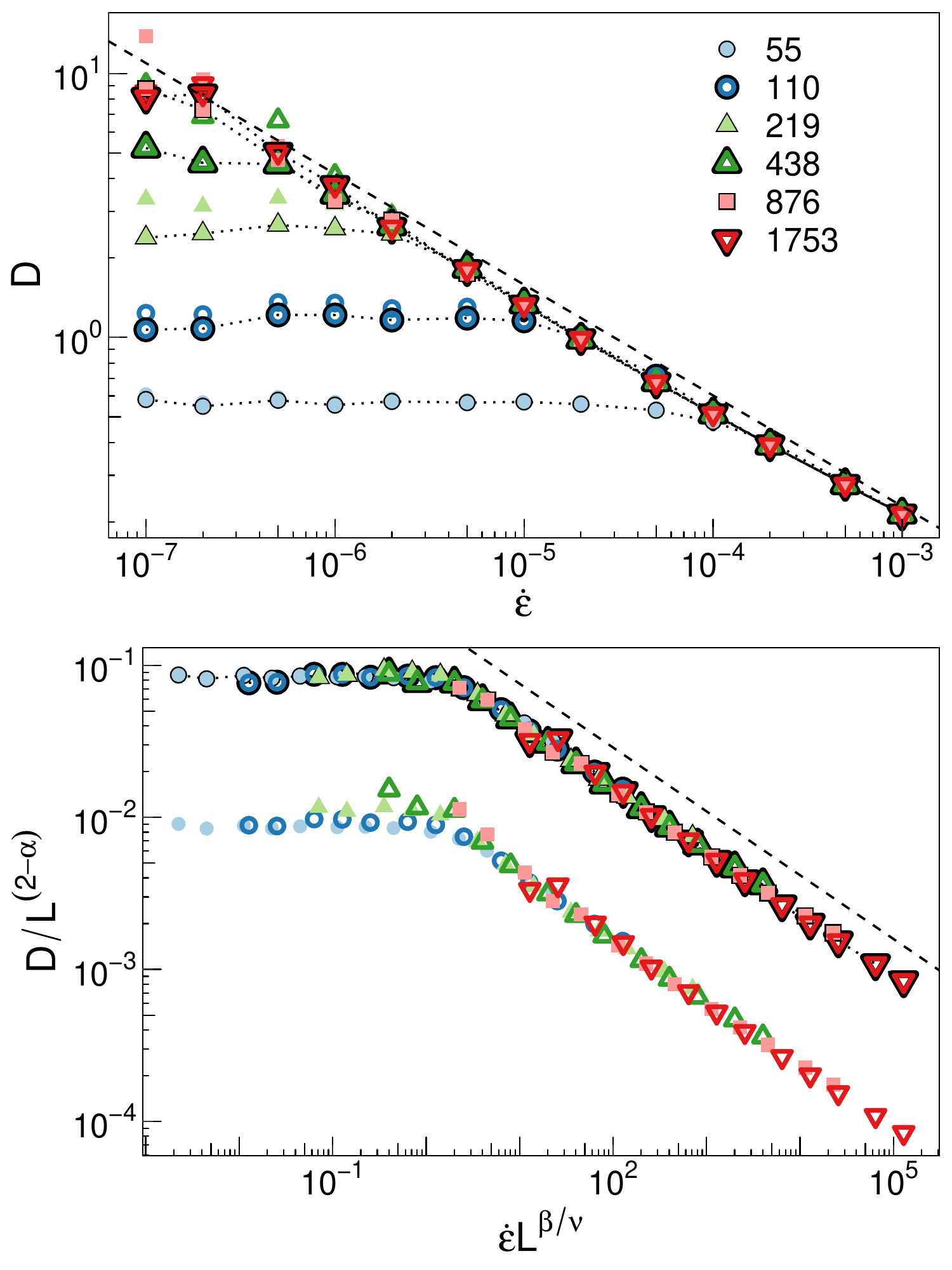}
	\caption{
	(a) The arithmetic and geometric average diffusion coefficients plotted as a function of rate for the indicated 2D system sizes. To distinguish the two metrics, the geometric average is joined by a black dotted line and there is a thin black outline around points. Note that the two sets of $L = 55$ data almost completely overlap. (b) Data is scaled according to the finite-size scaling procedure in Eq. \eqref{eq:diff_scaling} using a value of $\beta/\nu = 2.5$ and $d-\alpha = 1.05$. Data for the geometric average is shifted vertically by a factor of 10 for visibility. Dashed lines in both panels represent power-law scaling with an exponent $(\alpha-d)\nu/\beta = -0.42$.} \label{fig:diffusion2D}
	\end{center}
\end{figure}

To measure the overall diffusion of the system, we calculated both the arithmetic average, $D_\mathrm{AA}$, and the geometric (or logarithmic) average, $D_\mathrm{GA}$, of the diffusion coefficient across all strains. 
In Fig. \ref{fig:diffusion2D}(a), $D_\mathrm{AA}$ and $D_\mathrm{GA}$ are plotted as a function of $\dot{\epsilon}$ for the indicated $L$. 
At high rates, one cannot distinguish between the two measures.
Diffusion is also independent of $L$ and increases as a power of decreasing $\dot{\epsilon}$. 
The measured power law is consistent with an exponent of $(\alpha-d) \nu/\beta = -0.42$ from Eq. \eqref{eq:diffusion_fsr} using the values of $\alpha$, $\beta$, and $\nu$ reported in Table \ref{table:yield_exponents}.   
Note that in the FSR regime, the diffusion does not depend on $\epsilon_M$ implying this result is not influenced by the choice of boundary conditions.

At lower rates, the power-law divergence of $D$ is truncated at a rate that decreases with increasing $L$ for both measures. 
In the QS regime, $D_\mathrm{AA}$ plateaus at a higher value than $D_\mathrm{GA}$ due to the strong influence of fluctuations on the arithmetic average.
Note that a larger shift is seen for larger system sizes.
This is consistent with data in Fig. \ref{fig:diffusion_strain}(b) where fluctuations in diffusion with $\epsilon_M$ grow with increasing $L$. 

To capture this transition, we propose the following finite-size scaling ansatz:
\begin{equation}
D \sim L^{d-\alpha} f_D(\dot{\epsilon} L^{\beta/\nu}) \ \ ,
\label{eq:diff_scaling}
\end{equation}
where $f_D(x)$ is a universal scaling function. 
To satisfy Eq. \eqref{eq:diffusion_fsr} in the limit $x \gg 1$, $f_D(x) \sim x^{(\alpha-d)\nu/\beta}$.
To satisfy Eq. \eqref{eq:diffusion_scaling} in the limit $x \ll 1$, $f_D(x)$ approaches a constant. 
In Fig. \ref{fig:diffusion2D}(b), we use this theory to scale the measured diffusion coefficients using $\beta/\nu = 2.5$ and $d-\alpha = 1.05$. 
$D_\mathrm{AA}$, which is sensitive to fluctuations, collapses poorly but the more robust $D_\mathrm{GA}$ collapses very well.
The strength of the collapse independently confirms our value of $\beta/\nu$.
We also attempted to collapse the data in \ref{fig:diffusion2D}(a) using a more conventional scaling ansatz which assumed $D \sim L$ in the QS regime and, by analogy, $\xi$ in the FSR regime.
This alternate ansatz achieved a visibly poorer collapse.
This supports the theories proposed here and in Refs. \cite{Tyukodi2018, Tyukodi2019}.
However as $d-\alpha$ is very close to 1.0, we have insufficient data to definitively rule out the possibility that the diffusion actually scales as $L$ in the QS regime.

The strong variation in diffusion with the alignment of periodic boundary conditions is a novel finding for particle-based simulations and is not unique to the KR boundary conditions. 
A similar effect was identified in simple shear and conventional pure shear geometries. 
In simple shear, asymmetric nonaffine motion is seen between the flow and gradient directions as previously reported by Lema\^{i}tre and Caroli \cite{Lemaitre2007,Lemaitre2009}.
Particles generally exhibit large nonaffine displacements in the flow direction which do not grow linearly with strain or correspond to diffusive behavior.
However in the gradient direction, nonaffine displacement is diffusive. 
When the lattice vectors return to an orthogonal configuration, system-spanning avalanches aligned along the gradient direction perfectly remap across the periodic boundary.
When this occurs there is a similar increase in diffusion in the QS regime. 
Additionally for pure shear using conventional geometries, excess diffusion occurs when the ratio of the box lengths is an integer such that a $45^\circ$ line wraps back onto itself. 
We note that previous studies of diffusion were unlikely to see this effect due to small system sizes in simple shear \cite{Lemaitre2007, Lemaitre2009} or relatively small strain intervals in pure shear with conventional periodic boundaries \cite{Maloney2008}. 

\begin{figure}
\begin{center}
	\includegraphics[width=0.45\textwidth]{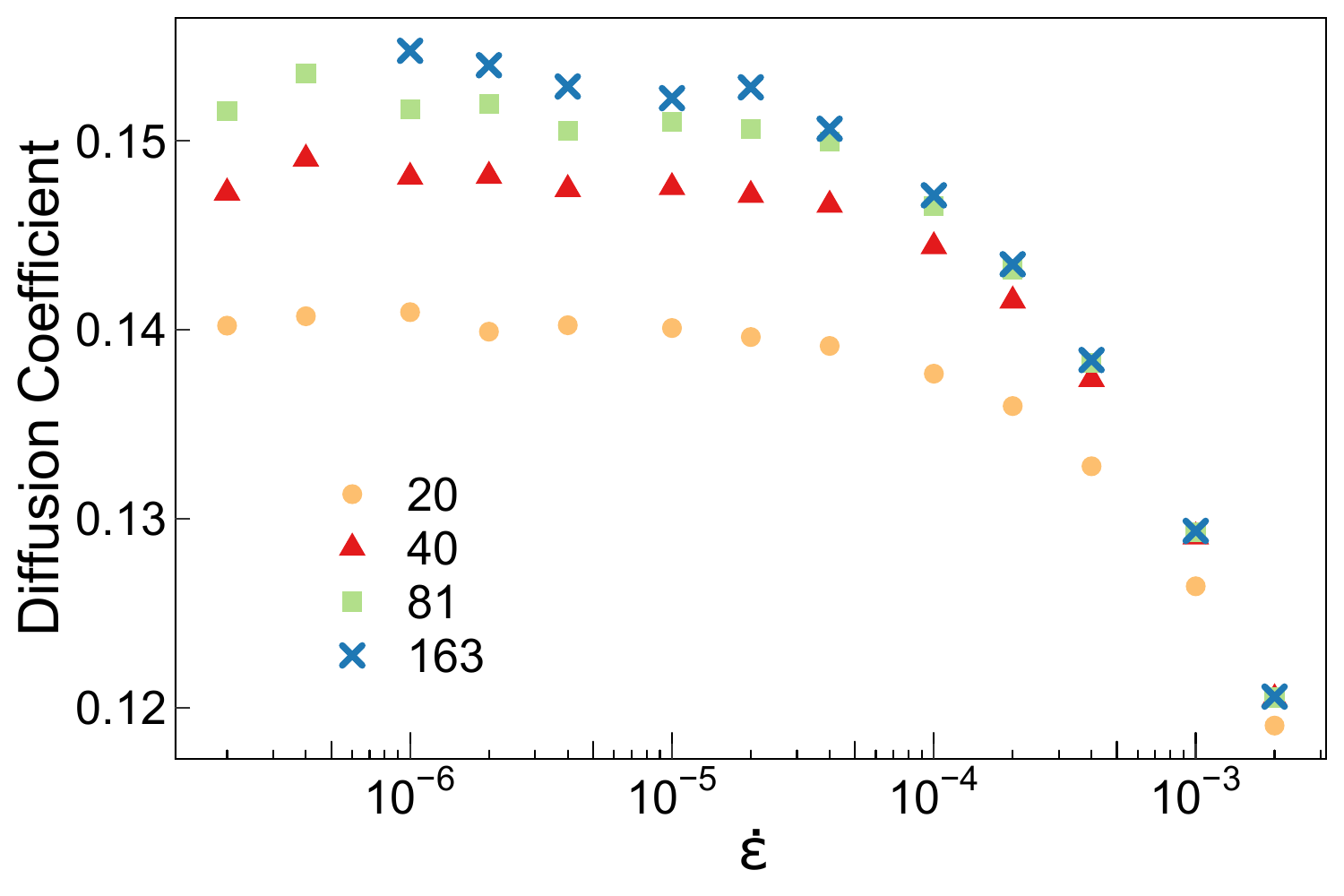}
	\caption{
	Diffusion coefficient as a function of strain rate for the system sizes indicated in the legend from 3D simulations.} \label{fig:diffusion3D}
	\end{center}
\end{figure}

In three dimensions, the diffusion is relatively simple. Previous QS studies failed to identify a strong dependence on system size \cite{SalernoThesis}. 
In Fig. \ref{fig:diffusion3D}, the diffusion coefficient is plotted as a function of strain rate for systems of size $L$ indicated in the legend. 
As the strain rate decreases, a small rise in diffusion can be identified before the diffusion saturates for all systems sizes below rates of $\sim 2\times 10^{-4}$. The plateau has a minor dependence on system size although it appears to reach an asymptotic maximum with increasing $L$. 
This suggests there is no divergence in diffusion at the critical point. 
In 3D, there exists a continuous range of possible slip planes with a varying azimuthal angle along which avalanches can grow. 
It is possible this scrambles any correlations in particle transport.

%%%%%%%%%%%%%%%%%%%%%%%%%%%%%%%%%%%%%%%%%%%%%%%%%%%%%%%%%%%%
\section{Summary and Conclusions}
\label{sec:yield_summary}
%%%%%%%%%%%%%%%%%%%%%%%%%%%%%%%%%%%%%%%%%%%%%%%%%%%%%%%%%%%%

Simulations of 2D and 3D sheared, disordered packings of LJ particles in steady state were used to identify critical exponents in the yielding transition. 
To reach steady-state flow, we employed KR and GKR boundary conditions.
These boundary conditions allowed simulations to reach large strains in pure shear without causing a reduction in one of the dimensions of the simulation cell.
This work focused on the effect of finite strain rates in order to identify scaling on the approach to the critical point. 
Exponents were accurately measured using finite-size scaling techniques including systems with over 4 million particles in 2D and over 7 million particles in 3D. 
The measured exponents are summarized in Table \ref{table:yield_exponents}.

\begin{center}
\begin{table}
\centering
\begin{tabular}{ |c|c|c|c|}
 \hline
  Values & 2D Estimates & 3D Estimates & Definition \\ 
 \hline
  $\alpha$ & $0.95 \pm 0.05$ & $1.15 \pm 0.05$ & $E_I \sim \ell_I^\alpha $ \\
	$\gamma$ & $1.3 \pm 0.1$ & $2.1 \pm 0.1$ & $R_{QS}(L,E) \sim L^\gamma$ \\
	$\tau$ & $1.3 \pm 0.1$ & $1.3 \pm 0.1$ & $R_{QS}(L,E) \sim E^{-\tau}$\\
  $\nu$ & $0.72 \pm 0.06$ & $0.48 \pm 0.06$ & $\xi \sim (\sigma - \sigma_c)^{-\nu}$ \\
  $\beta$ & $1.8 \pm 0.1$ & $1.5 \pm 0.05$ & $\dot{\epsilon} \sim (\sigma - \sigma_c)^\beta$ \\
  $z$ & $1.55 \pm 0.05$ & $1.25 \pm 0.05$ & $T_\mathrm{I} \sim \ell_\mathrm{I}^z$ \\
  $\phi$ & $1.0 \pm 0.1$ & $1.5 \pm 0.1$ & $\Delta \sigma \sim L^\phi$\\
  $x$ & $2.0 \pm 0.04$ & $3.0 \pm 0.08$ & $R_K \approx 1$ for $\dot{\epsilon} L^x \gg 1$\\
 \hline 
\end{tabular}
	\caption{Summary of critical exponents found here for 2D and 3D. The critical stress was $1.1897\pm 0.003$ in 2D and $1.2501 \pm 0.003$ in 3D. Values of $\gamma$ and $\tau$ are quoted from Refs. \cite{Salerno2012, Salerno2013}. Values of $\alpha$ and $z$ are quoted from Ref. \cite{Clemmer2021}.}
\label{table:yield_exponents}
\end{table}
\end{center}

At QS strain rates, the dynamics of the system are characterized by discrete, temporally-separated avalanches that are capable of spanning the system. 
With increasing rate, the maximum size of an avalanches transitions from being system-size limited to being limited by a rate-dependent correlation length $\xi \sim \dot{\epsilon}^{-\nu/\beta}$.
A scaling relation was proposed to provide a bound on $\beta/\nu$ and the transition to the FSR regime by considering the fraction of time that the system is active and undergoing plastic flow.
We also proposed the existence of a critical exponent $x$ that determines the emergence of quiescence in the system (Fig. \ref{krat}). 
This exponent is argued to equal $2\phi$ and provides an additional lower bound for $\beta/\nu$.

In the QS regime, the flow stress approaches a limiting value $\sigma_c$ with increasing system size. 
As rate increases, the average flow stress rises and finite-size effects disappear (Fig. \ref{average_stress}).
In the limit of infinite system size, the rise in stress grows as a power of rate with exponent $1/\beta$ (Fig. \ref{stress_beta}). 
Using a finite-size scaling ansatz, we collapsed curves of different system sizes and estimated values of $\beta$ and $\nu$ (Fig. \ref{stress_scaled}).
We note that our measured values of $\beta$ decrease with increasing spatial dimension.
This trend was noted in EPMs by Lin and Wyart, who postulated such a decrease could hint at a smaller mean-field value of $\beta = 1$ \cite{Lin2018}.

The exponent $\beta$ has been previously measured in MD studies which found $\beta = 2$ \cite{Chaudhuri2012} and $\beta = 2.33$ \cite{Karmakar2010a} in 2D and $\beta = 3$ \cite{Karmakar2010a} in 3D. 
These measurements were based on stress data from system sizes up to $\sim 10^4$ particles, equivalent to our smallest or second smallest system. 
These measurements, particularly in 3D, are not consistent with the values of $\beta$ found in this work.
However, one would expect a very narrow range of critical scaling for these small system sizes, possibly explaining the discrepancy. 

In this work, we also found fluctuations in the flow stress do not diverge with decreasing rate, but instead scale as $L^{-d/2}$ in all regimes (Fig. \ref{variance}).
This is consistent with previous MD results that demonstrated that $\phi = d/2$ \cite{Salerno2013}.  
This is further evidence that fluctuations in stress are set by the incoherent addition of $N\sim L^d$ incoherent signals. 
If the size of fluctuations were determined by the largest stress drops of avalanches, it would suggest that $\nu = 1/(d-\alpha)$ \cite{Salerno2013, Lin2014}. 
Based on our measurements of $\alpha$, this relation would predict a value of $\nu = 0.95 \pm 0.05$ and $0.54 \pm 0.05$ in 2D and 3D, respectively. 
We note that this scaling relation does not accurately predict $\nu$ in either dimension and would only be valid if $\nu = 1/\phi$. 
It has been argued that if $\phi = d/2$, as seen here, one does not expect to have an equality between $\nu$ and $1/\phi$ \cite{Pazmandi1997}.

Lin et al. similarly found that fluctuations in stress scaled as a power of $L$ with an exponent of approximately $0.86 \pm 0.03$ and $1.39 \pm 0.08$ in 2D and 3D EPMs, respectively \cite{Lin2014}. 
The authors argued this exponent was equivalent to $1/\nu$ implying $\nu = 1.16 \pm 0.04$ and $0.72 \pm 0.04$ in 2D and 3D, respectively. 
We note that neither measurement is consistent with our values of $\nu$ in Table \ref{table:yield_exponents}, however, these exponents are on the threshold of being consistent with our measurements of $\phi$ in 2D and 3D.

In 2D, the rate of particle diffusion was found to grow as a power of decreasing strain rate (Fig. \ref{fig:diffusion2D}).
In the QS regime, we identified that the diffusion depends on the geometry of the simulation box (Fig. \ref{fig:diffusion_strain}) for which we described a potential mechanism (Fig. \ref{fig:kr_grid}).
We proposed a rate and system-size dependent scaling theory based on arguments from Tyukodi et al. \cite{Tyukodi2018, Tyukodi2019} who suggested the diffusion in their 2D EPM will scale as $L^{2-\alpha}$ versus the commonly assumed $L$ in the QS regime due to the fractal nature of avalanches.
Our derivation suggests diffusion scales as $L^{d-\alpha}$ in the QS regime and $\xi^{d-\alpha}$ in the FSR regime.
Curves of the geometric averaged diffusion coefficient as a function of rate and system size were collapsed (Fig. \ref{fig:diffusion2D}b) providing a second measure of $\beta/\nu$.
Despite the very good data collapse, we could not conclusively rule out the possibility that the diffusion scales as $L$ in the QS regime due to the proximity of $\alpha$ to unity.
In 3D, we did not identify a divergence in diffusion with decreasing rate or increasing system size and postulated that this was due to an axial symmetry in the deformation geometry.

Finally, we note that a similar dependence of particle diffusion on system size and rate was previously identified by Lema\^{i}tre and Caroli in 2D MD simulations \cite{Lemaitre2009}. 
Data for different system sizes was collapsed using a finite-size scaling relation which assumed the diffusion scales as $L$ in the QS regime.
Notably, this collapse found a distinct value of $\beta/\nu = 2$. 
However, these simulations were also limited to smaller system sizes, up to $\sim 5\times10^5$ particles or smaller than our $L = 219$ system, which could explain the difference in our results.

EPM simulations have been successful at expanding our theoretical understanding of the yielding transition but there is an open question whether the yielding transitions in MD and EPM are in the same universality class.
Here we provide a brief comparison of exponents measured in EPMs to the exponents in Table \ref{table:yield_exponents}.
Avalanche distributions in EPMs are similar to those in MD, with measurements of $\tau$ of $1.25-1.36$ in 2D \cite{Talamali2011, Budrikis2013, Lin2014, Liu2016, Budrikis2017, Karimi2017, Tyukodi2019, Ferrero2019} and $1.25-1.45$ in 3D \cite{Lin2014, Liu2016, Budrikis2017}.
However, the range of values measured for $\alpha$ in EPMs, $0.9-1.1$ in 2D \cite{Lin2014, Liu2016, Karimi2017, Tyukodi2018, Tyukodi2019, Ferrero2019} and both $1.5 \pm 0.05$ \cite{Lin2014} and $1.3 \pm 0.1$ \cite{Liu2016} in 3D, have wide variability and are hard to compare to MD.
Only the smaller values are consistent with MD, although in 3D $\alpha$ is on the border of inconsistency.
As noted and explored by Ferrero and Jagla \cite{Ferrero2019}, it is important that future work identify whether the variability in exponents within EPMs is due to uncertainty in measurements and finite-size effects or whether different models can produce different avalanche statistics.
In the underdamped limit, distinct avalanche distributions are produced in both MD \cite{Salerno2012, Salerno2013} and finite-element based EPMs \cite{Karimi2017}, but evidence of critical behavior and finite-size scaling collapses are only found in MD.

The exponents $\beta$ and $z$ have also been measured in EPMs. 
Values for $\beta$ include $1.52 \pm 0.05$ \cite{Lin2014} and $1.54 \pm 0.02$ \cite{Liu2016} in 2D and $1.38 \pm 0.03$ \cite{Lin2014} and $1.55 \pm 0.02$ \cite{Liu2016} in 3D.
Work on 2D EPMs by Ferrero and Jagla found that $\beta$ can depend on dynamical rules, measuring values of both $3/2$ and $2$ \cite{Ferrero2019}.
All of these 2D values are distinct from our measurement, while in 3D only Ref. \cite{Liu2016} found a consistent value of $\beta$.
For the dynamic exponent, EPMs find $z < 1$ \cite{Lin2014, Liu2016, Ferrero2019} due instantaneous information propagation as noted in Refs. \cite{Lin2014, Lin2018}.
This is inherently inconsistent with MD which has finite speed of sound and a value of $z >1$, discussed further in the sibling paper \cite{Clemmer2021}.
These comparisons suggest that current EPMs may not be in the same dynamic critical universality class as MD.

In summary, this work has provided accurate measurement of exponents of the yielding transition using finite-size scaling in the overdamped limit.
Furthermore, this work provides new scaling theories to describe the critical behavior of yielding.
Such measurements and relations are important in determining the scope and nature of the dynamical critical point of the yielding transition.

\begin{acknowledgments}
The authors thank Craig Maloney for useful conversations. Calculations were performed at the Maryland Advanced Research Computing Center. This material is based upon work supported by the National Science Foundation under Grant No. DMR-1411144. MOR acknowledged support from the Simons Foundation.

The views and conclusions contained in this document are those of the authors and should not be interpreted as representing the official policies, either expressed or implied, of the Army Research Laboratory or the U.S. Government. The U.S. Government is authorized to reproduce and distribute reprints for Government purposes notwithstanding any copyright notation herein.

\end{acknowledgments}

%=====================================================================%
\appendix
%=====================================================================%
\beginappendix

%%%%%%%%%%%%%%%%%%%%%%%%%%%%%%%%%%%%%%%%%%%%%%%%%%%%%%%%%%%%
\section{Root Mean Square Kinetic Energy}
\label{sec:yield_rmske}
%%%%%%%%%%%%%%%%%%%%%%%%%%%%%%%%%%%%%%%%%%%%%%%%%%%%%%%%%%%%

\begin{figure}
\begin{center}
	\includegraphics[width=0.45\textwidth]{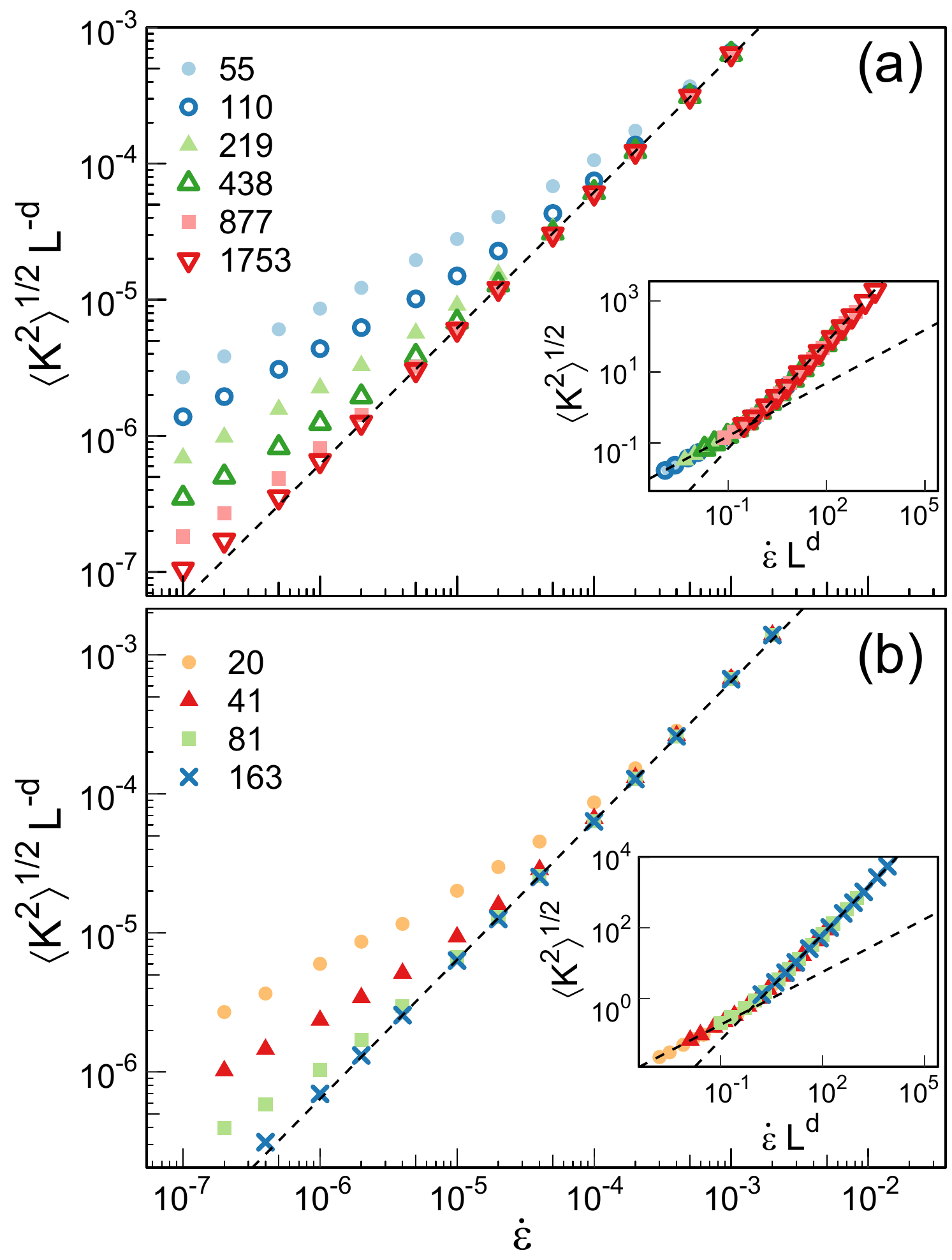}
	\caption{
		Root mean squared kinetic energy density $\sqrt{\langle K^2 \rangle}/L^d$ as a function of rate in (a) 2D and (b) 3D for the system sizes indicated in the legends.
	Dashed lines show power laws with exponents of unity.
		Insets show that the data collapses when rate and rms kinetic energy density are scaled by $L^x$ with $x = d$. Dashed lines in the insets indicate power laws with exponents of 1.0 and 0.5.
	}
	\label{k2}
	\end{center}
\end{figure}

In this appendix, we provide an alternate measurement of the exponent $x$ described in Sec. \ref{sec:transition} using the root mean square (rms) kinetic energy.
Figure \ref{k2} shows the variation of the rms kinetic energy density with rate for different $L$ in 2D and 3D.
In the QS regime, fluctuations in the kinetic energy are much larger than the average and $ \langle K^2 \rangle \sim L^{d} \dot{\epsilon}$ from Eq. \eqref{eq:var_ke}. 
In the FSR regime the fractional change in kinetic energy is small and$\sqrt{\langle K^2 \rangle } \approx \langle K \rangle \propto L^d \dot{\epsilon}$ where the final relation is from Eq. \eqref{eq:power}.
Thus the rms kinetic energy density should change from a square root to linear dependence on rate with increasing rate.
This behavior is clear in Fig. \ref{k2}.

The crossover to the limiting FSR behavior should occur when the fluctuations in $K$ are small compared to the mean.
This is just the condition that $R_K \equiv (K_\mathrm{max}-K_\mathrm{min})/K_\mathrm{max}$ is small. 
From above, $\sqrt{\langle K^2 \rangle} < \langle K \rangle$ will occur when $\dot{\epsilon} > L^{-d}$ implying $x = d$.
As shown in Fig. \ref{krat}, this occurs at a rate that scales as $L^{-x}$ with $x=d=2 \phi$.
The insets in Fig. \ref{k2} show that, as expected, the same scaling collapses results for the rms kinetic energy density. 
The collapses are consistent with estimates of $x = 2.00 \pm 0.04$ in 2D and $3.00 \pm 0.08$ in 3D.

%%%%%%%%%%%%%%%%%%%%%%%%%%%%%%%%%%%%%%%%%%%%%%%%%%%%%%%%%%%%
\section{Simple Shear Geometry}
\label{sec:yield_simple}
%%%%%%%%%%%%%%%%%%%%%%%%%%%%%%%%%%%%%%%%%%%%%%%%%%%%%%%%%%%%

Finally, we present data for simple shear deformation in  2D systems and discuss how results compare to results for pure shear with KR boundary conditions described in the main text.  
One fundamental distinction between these two methods is that the velocity gradient is always perpendicular to a periodic lattice vector of the unit cell in simple shear.
This allows shear to localize on bands that wrap around the periodic boundaries.
Simple shear also produces a local rotation that is absent in pure shear.

\begin{figure}
\begin{center}
	\includegraphics[width=0.45\textwidth]{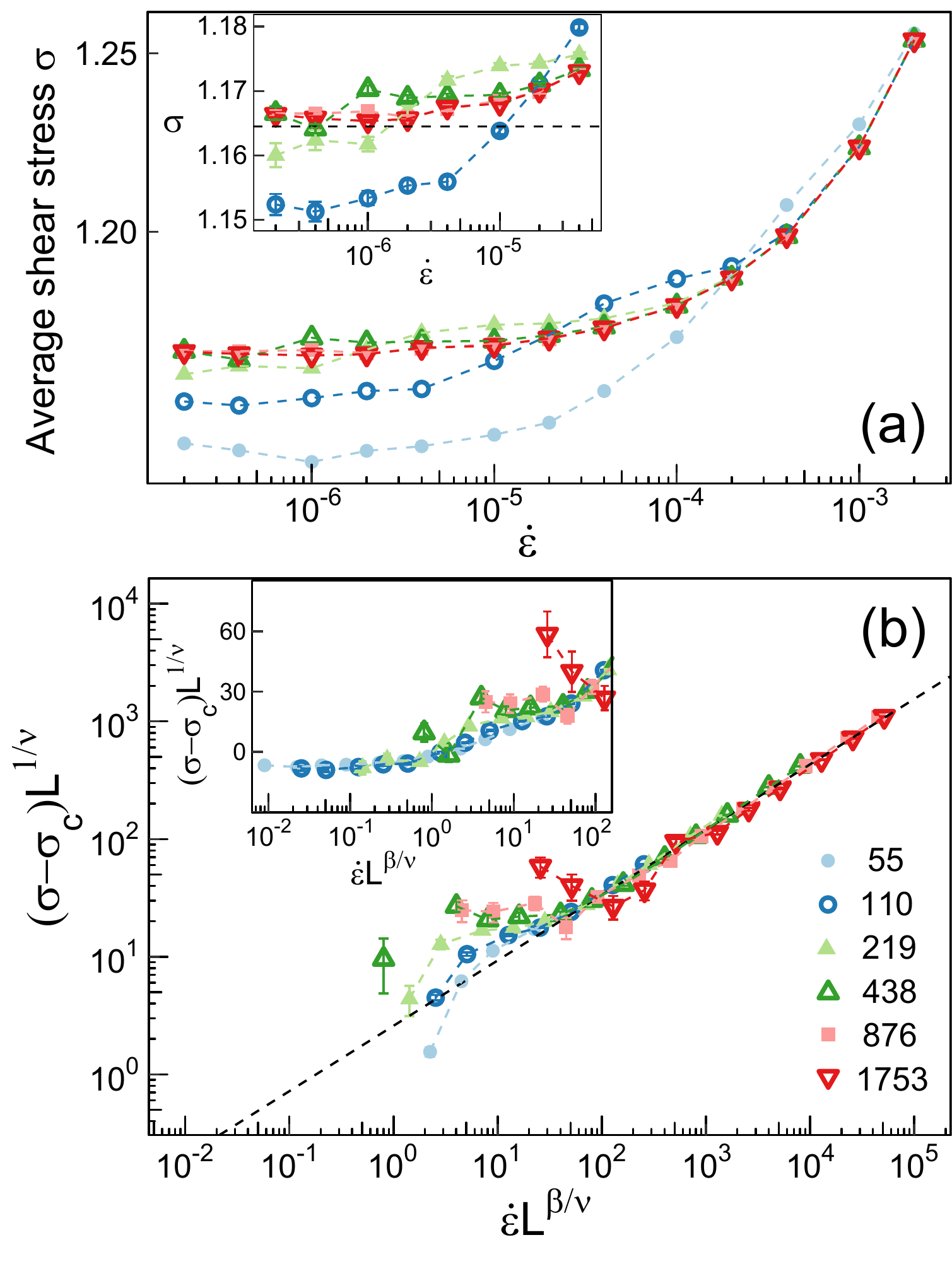}
	\caption{(a) The average shear stress as a function of strain rate during steady-state simple shear for 2D systems of size $L$ indicated in the legend. The inset contains an expanded view of low rate data. A dashed horizontal line highlights $\sigma_c = 1.1645$. (b) The above data rescaled according to the finite-size scaling relation in Eq. \eqref{eq:stress_scaling} using values of $\beta = 1.8$ and $\nu = 0.72$. The inset in (b) includes an expanded view of the same data using linear-log axes to highlight values of $\langle \sigma \rangle < \sigma_c$.} \label{fig:simple_stress}
	\end{center}
\end{figure}

In Fig. \ref{fig:simple_stress}(a), the average shear stress during simple shear deformation is plotted as a function of rate for 2D systems of size $L$ indicated in the legend. 
The trends in the data resemble those seen in Fig. \ref{average_stress}(a) except the onset of finite-size effects is marked by a shoulder in the shear stress. 
As system size decreases, the magnitude of the shoulder decreases and it moves to lower rates.

In Fig. \ref{fig:simple_stress}(b), the shear stress data is rescaled according to Eq. \eqref{eq:stress_scaling}. 
Notably in the high rate limit, the shear stress rises as a power of increasing strain rate with the expected value of $\beta$. 
Additionally, the curves for the three smallest systems collapse in the QS regime in the inset of Fig. \ref{fig:simple_stress}(b). 
We do not try to refine values of exponents or measure their error bars and only emphasize that the data is consistent with the values of $\beta$ and $\nu$ used for pure shear in the main text (Table \ref{table:yield_exponents}).
Additionally, $\sigma_c$ is estimated to be around $1.1645$, lower than $\sigma_c$ from pure shear. 
For simple shear, the critical stress cannot be accurately measured due to the previously mentioned shoulder in the shear stress. 

Due to the system-size dependent bump in $\langle \sigma \rangle$, the data in Fig. \ref{fig:simple_stress}(b) fails to collapse at intermediate rates. This suggests that while the critical exponents $\beta$ and $\nu$ do not depend on deformation geometry, the crossover function in Eq. \eqref{eq:stress_scaling} picks up an additional dependence on system size.
The onset of the bump in Fig. \ref{fig:simple_stress}(b) may scale as a power of $L$ with an exponent of $\beta/\nu$ but this cannot be verified due to error bars on data points and uncertainty in $\sigma_c$. 
Around this rate, avalanches have just begun to span the system and one can identify the presence of transient shear bands that cross the width of the system. 
In simple shear geometry, these shear bands always align with the periodic boundary conditions allowing for them to wrap back on to themselves, self-reinforcing their dynamics. 
This contrasts with the KR boundary conditions discussed in Sec. \ref{sec:yield_diffusion}. 
We therefore theorize that deviations in scaling seen for simple shear may be due to this unique geometry.

%=====================================================================%
\newpage
\bibliographystyle{apsrev4-1}

\begin{thebibliography}{66}%
\makeatletter
\providecommand \@ifxundefined [1]{%
 \@ifx{#1\undefined}
}%
\providecommand \@ifnum [1]{%
 \ifnum #1\expandafter \@firstoftwo
 \else \expandafter \@secondoftwo
 \fi
}%
\providecommand \@ifx [1]{%
 \ifx #1\expandafter \@firstoftwo
 \else \expandafter \@secondoftwo
 \fi
}%
\providecommand \natexlab [1]{#1}%
\providecommand \enquote  [1]{``#1''}%
\providecommand \bibnamefont  [1]{#1}%
\providecommand \bibfnamefont [1]{#1}%
\providecommand \citenamefont [1]{#1}%
\providecommand \href@noop [0]{\@secondoftwo}%
\providecommand \href [0]{\begingroup \@sanitize@url \@href}%
\providecommand \@href[1]{\@@startlink{#1}\@@href}%
\providecommand \@@href[1]{\endgroup#1\@@endlink}%
\providecommand \@sanitize@url [0]{\catcode `\\12\catcode `\$12\catcode
  `\&12\catcode `\#12\catcode `\^12\catcode `\_12\catcode `\%12\relax}%
\providecommand \@@startlink[1]{}%
\providecommand \@@endlink[0]{}%
\providecommand \url  [0]{\begingroup\@sanitize@url \@url }%
\providecommand \@url [1]{\endgroup\@href {#1}{\urlprefix }}%
\providecommand \urlprefix  [0]{URL }%
\providecommand \Eprint [0]{\href }%
\providecommand \doibase [0]{http://dx.doi.org/}%
\providecommand \selectlanguage [0]{\@gobble}%
\providecommand \bibinfo  [0]{\@secondoftwo}%
\providecommand \bibfield  [0]{\@secondoftwo}%
\providecommand \translation [1]{[#1]}%
\providecommand \BibitemOpen [0]{}%
\providecommand \bibitemStop [0]{}%
\providecommand \bibitemNoStop [0]{.\EOS\space}%
\providecommand \EOS [0]{\spacefactor3000\relax}%
\providecommand \BibitemShut  [1]{\csname bibitem#1\endcsname}%
\let\auto@bib@innerbib\@empty
%</preamble>
\bibitem [{\citenamefont {Bonn}\ \emph {et~al.}(2017)\citenamefont {Bonn},
  \citenamefont {Denn}, \citenamefont {Berthier}, \citenamefont {Divoux},\ and\
  \citenamefont {Manneville}}]{Bonn2017}%
  \BibitemOpen
  \bibfield  {author} {\bibinfo {author} {\bibfnamefont {D.}~\bibnamefont
  {Bonn}}, \bibinfo {author} {\bibfnamefont {M.~M.}\ \bibnamefont {Denn}},
  \bibinfo {author} {\bibfnamefont {L.}~\bibnamefont {Berthier}}, \bibinfo
  {author} {\bibfnamefont {T.}~\bibnamefont {Divoux}}, \ and\ \bibinfo {author}
  {\bibfnamefont {S.}~\bibnamefont {Manneville}},\ }\href {\doibase
  10.1103/RevModPhys.89.035005} {\bibfield  {journal} {\bibinfo  {journal}
  {Reviews of Modern Physics}\ }\textbf {\bibinfo {volume} {89}},\ \bibinfo
  {pages} {035005} (\bibinfo {year} {2017})}\BibitemShut {NoStop}%
\bibitem [{\citenamefont {Park}\ and\ \citenamefont {Durian}(1994)}]{Park1994}%
  \BibitemOpen
  \bibfield  {author} {\bibinfo {author} {\bibfnamefont {S.~S.}\ \bibnamefont
  {Park}}\ and\ \bibinfo {author} {\bibfnamefont {D.~J.}\ \bibnamefont
  {Durian}},\ }\href {\doibase 10.1103/PhysRevLett.72.3347} {\bibfield
  {journal} {\bibinfo  {journal} {Physical Review Letters}\ }\textbf {\bibinfo
  {volume} {72}},\ \bibinfo {pages} {3347} (\bibinfo {year}
  {1994})}\BibitemShut {NoStop}%
\bibitem [{\citenamefont {Durian}(1995)}]{Durian1995}%
  \BibitemOpen
  \bibfield  {author} {\bibinfo {author} {\bibfnamefont {D.~J.}\ \bibnamefont
  {Durian}},\ }\href {\doibase 10.1103/PhysRevLett.75.4780} {\bibfield
  {journal} {\bibinfo  {journal} {Physical Review Letters}\ }\textbf {\bibinfo
  {volume} {75}},\ \bibinfo {pages} {4780} (\bibinfo {year}
  {1995})}\BibitemShut {NoStop}%
\bibitem [{\citenamefont {Dennin}(2004)}]{Dennin2004}%
  \BibitemOpen
  \bibfield  {author} {\bibinfo {author} {\bibfnamefont {M.}~\bibnamefont
  {Dennin}},\ }\href {\doibase 10.1103/PhysRevE.70.041406} {\bibfield
  {journal} {\bibinfo  {journal} {Physical Review E}\ }\textbf {\bibinfo
  {volume} {70}},\ \bibinfo {pages} {041406} (\bibinfo {year} {2004})},\
  \Eprint {http://arxiv.org/abs/0405489} {0405489 [cond-mat]} \BibitemShut
  {NoStop}%
\bibitem [{\citenamefont {Mason}\ \emph {et~al.}(1996)\citenamefont {Mason},
  \citenamefont {Bibette},\ and\ \citenamefont {Weitz}}]{Mason1996}%
  \BibitemOpen
  \bibfield  {author} {\bibinfo {author} {\bibfnamefont {T.}~\bibnamefont
  {Mason}}, \bibinfo {author} {\bibfnamefont {J.}~\bibnamefont {Bibette}}, \
  and\ \bibinfo {author} {\bibfnamefont {D.}~\bibnamefont {Weitz}},\ }\href
  {\doibase 10.1006/jcis.1996.0235} {\bibfield  {journal} {\bibinfo  {journal}
  {Journal of Colloid and Interface Science}\ }\textbf {\bibinfo {volume}
  {179}},\ \bibinfo {pages} {439} (\bibinfo {year} {1996})}\BibitemShut
  {NoStop}%
\bibitem [{\citenamefont {Coussot}\ \emph {et~al.}(2002)\citenamefont
  {Coussot}, \citenamefont {Nguyen}, \citenamefont {Huynh},\ and\ \citenamefont
  {Bonn}}]{Coussot2002}%
  \BibitemOpen
  \bibfield  {author} {\bibinfo {author} {\bibfnamefont {P.}~\bibnamefont
  {Coussot}}, \bibinfo {author} {\bibfnamefont {Q.~D.}\ \bibnamefont {Nguyen}},
  \bibinfo {author} {\bibfnamefont {H.~T.}\ \bibnamefont {Huynh}}, \ and\
  \bibinfo {author} {\bibfnamefont {D.}~\bibnamefont {Bonn}},\ }\href {\doibase
  10.1103/PhysRevLett.88.175501} {\bibfield  {journal} {\bibinfo  {journal}
  {Physical Review Letters}\ }\textbf {\bibinfo {volume} {88}},\ \bibinfo
  {pages} {175501} (\bibinfo {year} {2002})}\BibitemShut {NoStop}%
\bibitem [{\citenamefont {Miller}\ \emph {et~al.}(1996)\citenamefont {Miller},
  \citenamefont {O'Hern},\ and\ \citenamefont {Behringer}}]{Miller1996}%
  \BibitemOpen
  \bibfield  {author} {\bibinfo {author} {\bibfnamefont {B.}~\bibnamefont
  {Miller}}, \bibinfo {author} {\bibfnamefont {C.}~\bibnamefont {O'Hern}}, \
  and\ \bibinfo {author} {\bibfnamefont {R.~P.}\ \bibnamefont {Behringer}},\
  }\href {\doibase 10.1103/PhysRevLett.77.3110} {\bibfield  {journal} {\bibinfo
   {journal} {Physical Review Letters}\ }\textbf {\bibinfo {volume} {77}},\
  \bibinfo {pages} {3110} (\bibinfo {year} {1996})}\BibitemShut {NoStop}%
\bibitem [{\citenamefont {Hayman}\ \emph {et~al.}(2011)\citenamefont {Hayman},
  \citenamefont {Duclou{\'{e}}}, \citenamefont {Foco},\ and\ \citenamefont
  {Daniels}}]{Hayman2011}%
  \BibitemOpen
  \bibfield  {author} {\bibinfo {author} {\bibfnamefont {N.~W.}\ \bibnamefont
  {Hayman}}, \bibinfo {author} {\bibfnamefont {L.}~\bibnamefont
  {Duclou{\'{e}}}}, \bibinfo {author} {\bibfnamefont {K.~L.}\ \bibnamefont
  {Foco}}, \ and\ \bibinfo {author} {\bibfnamefont {K.~E.}\ \bibnamefont
  {Daniels}},\ }\href {\doibase 10.1007/s00024-011-0269-3} {\bibfield
  {journal} {\bibinfo  {journal} {Pure and Applied Geophysics}\ }\textbf
  {\bibinfo {volume} {168}},\ \bibinfo {pages} {2239} (\bibinfo {year}
  {2011})}\BibitemShut {NoStop}%
\bibitem [{\citenamefont {Sun}\ \emph {et~al.}(2010)\citenamefont {Sun},
  \citenamefont {Yu}, \citenamefont {Jiao}, \citenamefont {Bai}, \citenamefont
  {Zhao},\ and\ \citenamefont {Wang}}]{Sun2010}%
  \BibitemOpen
  \bibfield  {author} {\bibinfo {author} {\bibfnamefont {B.~a.}\ \bibnamefont
  {Sun}}, \bibinfo {author} {\bibfnamefont {H.~B.}\ \bibnamefont {Yu}},
  \bibinfo {author} {\bibfnamefont {W.}~\bibnamefont {Jiao}}, \bibinfo {author}
  {\bibfnamefont {H.~Y.}\ \bibnamefont {Bai}}, \bibinfo {author} {\bibfnamefont
  {D.~Q.}\ \bibnamefont {Zhao}}, \ and\ \bibinfo {author} {\bibfnamefont
  {W.~H.}\ \bibnamefont {Wang}},\ }\href {\doibase
  10.1103/PhysRevLett.105.035501} {\bibfield  {journal} {\bibinfo  {journal}
  {Physical Review Letters}\ }\textbf {\bibinfo {volume} {105}},\ \bibinfo
  {pages} {035501} (\bibinfo {year} {2010})}\BibitemShut {NoStop}%
\bibitem [{\citenamefont {Sun}\ \emph {et~al.}(2012)\citenamefont {Sun},
  \citenamefont {Pauly}, \citenamefont {Tan}, \citenamefont {Stoica},
  \citenamefont {Wang}, \citenamefont {K{\"{u}}hn},\ and\ \citenamefont
  {Eckert}}]{Sun2012}%
  \BibitemOpen
  \bibfield  {author} {\bibinfo {author} {\bibfnamefont {B.~A.}\ \bibnamefont
  {Sun}}, \bibinfo {author} {\bibfnamefont {S.}~\bibnamefont {Pauly}}, \bibinfo
  {author} {\bibfnamefont {J.}~\bibnamefont {Tan}}, \bibinfo {author}
  {\bibfnamefont {M.}~\bibnamefont {Stoica}}, \bibinfo {author} {\bibfnamefont
  {W.~H.}\ \bibnamefont {Wang}}, \bibinfo {author} {\bibfnamefont
  {U.}~\bibnamefont {K{\"{u}}hn}}, \ and\ \bibinfo {author} {\bibfnamefont
  {J.}~\bibnamefont {Eckert}},\ }\href {\doibase 10.1016/j.actamat.2012.04.013}
  {\bibfield  {journal} {\bibinfo  {journal} {Acta Materialia}\ }\textbf
  {\bibinfo {volume} {60}},\ \bibinfo {pages} {4160} (\bibinfo {year}
  {2012})}\BibitemShut {NoStop}%
\bibitem [{\citenamefont {Antonaglia}\ \emph {et~al.}(2014)\citenamefont
  {Antonaglia}, \citenamefont {Wright}, \citenamefont {Gu}, \citenamefont
  {Byer}, \citenamefont {Hufnagel}, \citenamefont {LeBlanc}, \citenamefont
  {Uhl},\ and\ \citenamefont {Dahmen}}]{Antonaglia2014}%
  \BibitemOpen
  \bibfield  {author} {\bibinfo {author} {\bibfnamefont {J.}~\bibnamefont
  {Antonaglia}}, \bibinfo {author} {\bibfnamefont {W.~J.}\ \bibnamefont
  {Wright}}, \bibinfo {author} {\bibfnamefont {X.}~\bibnamefont {Gu}}, \bibinfo
  {author} {\bibfnamefont {R.~R.}\ \bibnamefont {Byer}}, \bibinfo {author}
  {\bibfnamefont {T.~C.}\ \bibnamefont {Hufnagel}}, \bibinfo {author}
  {\bibfnamefont {M.}~\bibnamefont {LeBlanc}}, \bibinfo {author} {\bibfnamefont
  {J.~T.}\ \bibnamefont {Uhl}}, \ and\ \bibinfo {author} {\bibfnamefont
  {K.~a.}\ \bibnamefont {Dahmen}},\ }\href {\doibase
  10.1103/PhysRevLett.112.155501} {\bibfield  {journal} {\bibinfo  {journal}
  {Physical Review Letters}\ }\textbf {\bibinfo {volume} {112}},\ \bibinfo
  {pages} {155501} (\bibinfo {year} {2014})}\BibitemShut {NoStop}%
\bibitem [{\citenamefont {Lin}\ \emph {et~al.}(2014{\natexlab{a}})\citenamefont
  {Lin}, \citenamefont {Lerner}, \citenamefont {Rosso},\ and\ \citenamefont
  {Wyart}}]{Lin2014}%
  \BibitemOpen
  \bibfield  {author} {\bibinfo {author} {\bibfnamefont {J.}~\bibnamefont
  {Lin}}, \bibinfo {author} {\bibfnamefont {E.}~\bibnamefont {Lerner}},
  \bibinfo {author} {\bibfnamefont {A.}~\bibnamefont {Rosso}}, \ and\ \bibinfo
  {author} {\bibfnamefont {M.}~\bibnamefont {Wyart}},\ }\href {\doibase
  10.1073/pnas.1406391111} {\bibfield  {journal} {\bibinfo  {journal}
  {Proceedings of the National Academy of Sciences}\ }\textbf {\bibinfo
  {volume} {111}},\ \bibinfo {pages} {14382} (\bibinfo {year}
  {2014}{\natexlab{a}})}\BibitemShut {NoStop}%
\bibitem [{\citenamefont {Liu}\ and\ \citenamefont {Nagel}(2010)}]{Liu2010}%
  \BibitemOpen
  \bibfield  {author} {\bibinfo {author} {\bibfnamefont {A.~J.}\ \bibnamefont
  {Liu}}\ and\ \bibinfo {author} {\bibfnamefont {S.~R.}\ \bibnamefont
  {Nagel}},\ }\href {\doibase 10.1146/annurev-conmatphys-070909-104045}
  {\bibfield  {journal} {\bibinfo  {journal} {Annual Review of Condensed Matter
  Physics}\ }\textbf {\bibinfo {volume} {1}},\ \bibinfo {pages} {347} (\bibinfo
  {year} {2010})}\BibitemShut {NoStop}%
\bibitem [{\citenamefont {Fisher}(1998)}]{Fisher1998}%
  \BibitemOpen
  \bibfield  {author} {\bibinfo {author} {\bibfnamefont {D.~S.}\ \bibnamefont
  {Fisher}},\ }\href {\doibase 10.1016/S0370-1573(98)00008-8} {\bibfield
  {journal} {\bibinfo  {journal} {Physics Reports}\ }\textbf {\bibinfo {volume}
  {301}},\ \bibinfo {pages} {113} (\bibinfo {year} {1998})}\BibitemShut
  {NoStop}%
\bibitem [{\citenamefont {Ji}\ and\ \citenamefont {Robbins}(1991)}]{Ji1991}%
  \BibitemOpen
  \bibfield  {author} {\bibinfo {author} {\bibfnamefont {H.}~\bibnamefont
  {Ji}}\ and\ \bibinfo {author} {\bibfnamefont {M.~O.}\ \bibnamefont
  {Robbins}},\ }\href {\doibase 10.1103/PhysRevA.44.2538} {\bibfield  {journal}
  {\bibinfo  {journal} {Physical Review A}\ }\textbf {\bibinfo {volume} {44}},\
  \bibinfo {pages} {2538} (\bibinfo {year} {1991})}\BibitemShut {NoStop}%
\bibitem [{\citenamefont {Martys}\ \emph
  {et~al.}(1991{\natexlab{a}})\citenamefont {Martys}, \citenamefont {Cieplak},\
  and\ \citenamefont {Robbins}}]{Martys1991}%
  \BibitemOpen
  \bibfield  {author} {\bibinfo {author} {\bibfnamefont {N.}~\bibnamefont
  {Martys}}, \bibinfo {author} {\bibfnamefont {M.}~\bibnamefont {Cieplak}}, \
  and\ \bibinfo {author} {\bibfnamefont {M.~O.}\ \bibnamefont {Robbins}},\
  }\href {\doibase 10.1103/PhysRevLett.66.1058} {\bibfield  {journal} {\bibinfo
   {journal} {Physical Review Letters}\ }\textbf {\bibinfo {volume} {66}},\
  \bibinfo {pages} {1058} (\bibinfo {year} {1991}{\natexlab{a}})}\BibitemShut
  {NoStop}%
\bibitem [{\citenamefont {M{\aa}l{\o}y}\ \emph {et~al.}(2006)\citenamefont
  {M{\aa}l{\o}y}, \citenamefont {Santucci}, \citenamefont {Schmittbuhl},\ and\
  \citenamefont {Toussaint}}]{Maloy2006}%
  \BibitemOpen
  \bibfield  {author} {\bibinfo {author} {\bibfnamefont {K.~J.}\ \bibnamefont
  {M{\aa}l{\o}y}}, \bibinfo {author} {\bibfnamefont {S.}~\bibnamefont
  {Santucci}}, \bibinfo {author} {\bibfnamefont {J.}~\bibnamefont
  {Schmittbuhl}}, \ and\ \bibinfo {author} {\bibfnamefont {R.}~\bibnamefont
  {Toussaint}},\ }\href {\doibase 10.1103/PhysRevLett.96.045501} {\bibfield
  {journal} {\bibinfo  {journal} {Physical Review Letters}\ }\textbf {\bibinfo
  {volume} {96}},\ \bibinfo {pages} {045501} (\bibinfo {year}
  {2006})}\BibitemShut {NoStop}%
\bibitem [{\citenamefont {Herschel}\ and\ \citenamefont
  {Bulkley}(1926)}]{Herschel1926}%
  \BibitemOpen
  \bibfield  {author} {\bibinfo {author} {\bibfnamefont {W.~H.}\ \bibnamefont
  {Herschel}}\ and\ \bibinfo {author} {\bibfnamefont {R.}~\bibnamefont
  {Bulkley}},\ }\href {\doibase 10.1007/BF01432034} {\bibfield  {journal}
  {\bibinfo  {journal} {Kolloid-Zeitschrift}\ }\textbf {\bibinfo {volume}
  {39}},\ \bibinfo {pages} {291} (\bibinfo {year} {1926})}\BibitemShut
  {NoStop}%
\bibitem [{\citenamefont {Tong}\ \emph {et~al.}(2016)\citenamefont {Tong},
  \citenamefont {Wang}, \citenamefont {Yi}, \citenamefont {Ren}, \citenamefont
  {Pauly}, \citenamefont {Gao}, \citenamefont {Zhai}, \citenamefont {Mattern},
  \citenamefont {Dahmen}, \citenamefont {Liaw},\ and\ \citenamefont
  {Eckert}}]{Tong2016}%
  \BibitemOpen
  \bibfield  {author} {\bibinfo {author} {\bibfnamefont {X.}~\bibnamefont
  {Tong}}, \bibinfo {author} {\bibfnamefont {G.}~\bibnamefont {Wang}}, \bibinfo
  {author} {\bibfnamefont {J.}~\bibnamefont {Yi}}, \bibinfo {author}
  {\bibfnamefont {J.~L.}\ \bibnamefont {Ren}}, \bibinfo {author} {\bibfnamefont
  {S.}~\bibnamefont {Pauly}}, \bibinfo {author} {\bibfnamefont {Y.~L.}\
  \bibnamefont {Gao}}, \bibinfo {author} {\bibfnamefont {Q.~J.}\ \bibnamefont
  {Zhai}}, \bibinfo {author} {\bibfnamefont {N.}~\bibnamefont {Mattern}},
  \bibinfo {author} {\bibfnamefont {K.~A.}\ \bibnamefont {Dahmen}}, \bibinfo
  {author} {\bibfnamefont {P.~K.}\ \bibnamefont {Liaw}}, \ and\ \bibinfo
  {author} {\bibfnamefont {J.}~\bibnamefont {Eckert}},\ }\href {\doibase
  10.1016/j.ijplas.2015.10.006} {\bibfield  {journal} {\bibinfo  {journal}
  {International Journal of Plasticity}\ }\textbf {\bibinfo {volume} {77}},\
  \bibinfo {pages} {141} (\bibinfo {year} {2016})}\BibitemShut {NoStop}%
\bibitem [{\citenamefont {Denisov}\ \emph {et~al.}(2016)\citenamefont
  {Denisov}, \citenamefont {L{\"{o}}rincz}, \citenamefont {Uhl}, \citenamefont
  {Dahmen},\ and\ \citenamefont {Schall}}]{Denisov2016}%
  \BibitemOpen
  \bibfield  {author} {\bibinfo {author} {\bibfnamefont {D.~V.}\ \bibnamefont
  {Denisov}}, \bibinfo {author} {\bibfnamefont {K.~A.}\ \bibnamefont
  {L{\"{o}}rincz}}, \bibinfo {author} {\bibfnamefont {J.~T.}\ \bibnamefont
  {Uhl}}, \bibinfo {author} {\bibfnamefont {K.~A.}\ \bibnamefont {Dahmen}}, \
  and\ \bibinfo {author} {\bibfnamefont {P.}~\bibnamefont {Schall}},\ }\href
  {\doibase 10.1038/ncomms10641} {\bibfield  {journal} {\bibinfo  {journal}
  {Nature Communications}\ }\textbf {\bibinfo {volume} {7}} (\bibinfo {year}
  {2016}),\ 10.1038/ncomms10641}\BibitemShut {NoStop}%
\bibitem [{\citenamefont {Bar{\'{e}}s}\ \emph {et~al.}(2017)\citenamefont
  {Bar{\'{e}}s}, \citenamefont {Wang}, \citenamefont {Wang}, \citenamefont
  {Bertrand}, \citenamefont {O'Hern},\ and\ \citenamefont
  {Behringer}}]{Bares2017}%
  \BibitemOpen
  \bibfield  {author} {\bibinfo {author} {\bibfnamefont {J.}~\bibnamefont
  {Bar{\'{e}}s}}, \bibinfo {author} {\bibfnamefont {D.}~\bibnamefont {Wang}},
  \bibinfo {author} {\bibfnamefont {D.}~\bibnamefont {Wang}}, \bibinfo {author}
  {\bibfnamefont {T.}~\bibnamefont {Bertrand}}, \bibinfo {author}
  {\bibfnamefont {C.~S.}\ \bibnamefont {O'Hern}}, \ and\ \bibinfo {author}
  {\bibfnamefont {R.~P.}\ \bibnamefont {Behringer}},\ }\href@noop {} {\bibfield
   {journal} {\bibinfo  {journal} {Physical Review E}\ }\textbf {\bibinfo
  {volume} {96}},\ \bibinfo {pages} {1} (\bibinfo {year} {2017})}\BibitemShut
  {NoStop}%
\bibitem [{\citenamefont {Salerno}\ \emph {et~al.}(2012)\citenamefont
  {Salerno}, \citenamefont {Maloney},\ and\ \citenamefont
  {Robbins}}]{Salerno2012}%
  \BibitemOpen
  \bibfield  {author} {\bibinfo {author} {\bibfnamefont {K.~M.}\ \bibnamefont
  {Salerno}}, \bibinfo {author} {\bibfnamefont {C.~E.}\ \bibnamefont
  {Maloney}}, \ and\ \bibinfo {author} {\bibfnamefont {M.~O.}\ \bibnamefont
  {Robbins}},\ }\href {\doibase 10.1103/PhysRevLett.109.105703} {\bibfield
  {journal} {\bibinfo  {journal} {Physical Review Letters}\ }\textbf {\bibinfo
  {volume} {109}},\ \bibinfo {pages} {105703} (\bibinfo {year}
  {2012})}\BibitemShut {NoStop}%
\bibitem [{\citenamefont {Salerno}\ and\ \citenamefont
  {Robbins}(2013)}]{Salerno2013}%
  \BibitemOpen
  \bibfield  {author} {\bibinfo {author} {\bibfnamefont {K.~M.}\ \bibnamefont
  {Salerno}}\ and\ \bibinfo {author} {\bibfnamefont {M.~O.}\ \bibnamefont
  {Robbins}},\ }\href {\doibase 10.1103/PhysRevE.88.062206} {\bibfield
  {journal} {\bibinfo  {journal} {Physical Review E}\ }\textbf {\bibinfo
  {volume} {88}},\ \bibinfo {pages} {062206} (\bibinfo {year}
  {2013})}\BibitemShut {NoStop}%
\bibitem [{\citenamefont {Nicolas}\ \emph {et~al.}(2018)\citenamefont
  {Nicolas}, \citenamefont {Ferrero}, \citenamefont {Martens},\ and\
  \citenamefont {Barrat}}]{Nicolas2018}%
  \BibitemOpen
  \bibfield  {author} {\bibinfo {author} {\bibfnamefont {A.}~\bibnamefont
  {Nicolas}}, \bibinfo {author} {\bibfnamefont {E.~E.}\ \bibnamefont
  {Ferrero}}, \bibinfo {author} {\bibfnamefont {K.}~\bibnamefont {Martens}}, \
  and\ \bibinfo {author} {\bibfnamefont {J.~L.}\ \bibnamefont {Barrat}},\
  }\href {\doibase 10.1103/RevModPhys.90.045006} {\bibfield  {journal}
  {\bibinfo  {journal} {Reviews of Modern Physics}\ }\textbf {\bibinfo {volume}
  {90}},\ \bibinfo {pages} {45006} (\bibinfo {year} {2018})}\BibitemShut
  {NoStop}%
\bibitem [{\citenamefont {Talamali}\ \emph {et~al.}(2011)\citenamefont
  {Talamali}, \citenamefont {Pet{\"{a}}j{\"{a}}}, \citenamefont
  {Vandembroucq},\ and\ \citenamefont {Roux}}]{Talamali2011}%
  \BibitemOpen
  \bibfield  {author} {\bibinfo {author} {\bibfnamefont {M.}~\bibnamefont
  {Talamali}}, \bibinfo {author} {\bibfnamefont {V.}~\bibnamefont
  {Pet{\"{a}}j{\"{a}}}}, \bibinfo {author} {\bibfnamefont {D.}~\bibnamefont
  {Vandembroucq}}, \ and\ \bibinfo {author} {\bibfnamefont {S.}~\bibnamefont
  {Roux}},\ }\href {\doibase 10.1103/PhysRevE.84.016115} {\bibfield  {journal}
  {\bibinfo  {journal} {Physical Review E}\ }\textbf {\bibinfo {volume} {84}},\
  \bibinfo {pages} {016115} (\bibinfo {year} {2011})},\ \Eprint
  {http://arxiv.org/abs/1103.5017} {1103.5017} \BibitemShut {NoStop}%
\bibitem [{\citenamefont {Budrikis}\ and\ \citenamefont
  {Zapperi}(2013)}]{Budrikis2013}%
  \BibitemOpen
  \bibfield  {author} {\bibinfo {author} {\bibfnamefont {Z.}~\bibnamefont
  {Budrikis}}\ and\ \bibinfo {author} {\bibfnamefont {S.}~\bibnamefont
  {Zapperi}},\ }\href {\doibase 10.1103/PhysRevE.88.062403} {\bibfield
  {journal} {\bibinfo  {journal} {Physical Review E}\ }\textbf {\bibinfo
  {volume} {88}},\ \bibinfo {pages} {062403} (\bibinfo {year}
  {2013})}\BibitemShut {NoStop}%
\bibitem [{\citenamefont {Lin}\ \emph {et~al.}(2014{\natexlab{b}})\citenamefont
  {Lin}, \citenamefont {Saade}, \citenamefont {Lerner}, \citenamefont {Rosso},\
  and\ \citenamefont {Wyart}}]{Lin2014a}%
  \BibitemOpen
  \bibfield  {author} {\bibinfo {author} {\bibfnamefont {J.}~\bibnamefont
  {Lin}}, \bibinfo {author} {\bibfnamefont {A.}~\bibnamefont {Saade}}, \bibinfo
  {author} {\bibfnamefont {E.}~\bibnamefont {Lerner}}, \bibinfo {author}
  {\bibfnamefont {A.}~\bibnamefont {Rosso}}, \ and\ \bibinfo {author}
  {\bibfnamefont {M.}~\bibnamefont {Wyart}},\ }\href {\doibase
  10.1209/0295-5075/105/26003} {\bibfield  {journal} {\bibinfo  {journal} {EPL
  (Europhysics Letters)}\ }\textbf {\bibinfo {volume} {105}},\ \bibinfo {pages}
  {26003} (\bibinfo {year} {2014}{\natexlab{b}})},\ \Eprint
  {http://arxiv.org/abs/1307.1646} {1307.1646} \BibitemShut {NoStop}%
\bibitem [{\citenamefont {Liu}\ \emph {et~al.}(2016)\citenamefont {Liu},
  \citenamefont {Ferrero}, \citenamefont {Puosi}, \citenamefont {Barrat},\ and\
  \citenamefont {Martens}}]{Liu2016}%
  \BibitemOpen
  \bibfield  {author} {\bibinfo {author} {\bibfnamefont {C.}~\bibnamefont
  {Liu}}, \bibinfo {author} {\bibfnamefont {E.~E.}\ \bibnamefont {Ferrero}},
  \bibinfo {author} {\bibfnamefont {F.}~\bibnamefont {Puosi}}, \bibinfo
  {author} {\bibfnamefont {J.-l.}\ \bibnamefont {Barrat}}, \ and\ \bibinfo
  {author} {\bibfnamefont {K.}~\bibnamefont {Martens}},\ }\href {\doibase
  10.1103/PhysRevLett.116.065501} {\bibfield  {journal} {\bibinfo  {journal}
  {Physical Review Letters}\ }\textbf {\bibinfo {volume} {116}},\ \bibinfo
  {pages} {065501} (\bibinfo {year} {2016})},\ \Eprint
  {http://arxiv.org/abs/1506.08161} {1506.08161} \BibitemShut {NoStop}%
\bibitem [{\citenamefont {Budrikis}\ \emph {et~al.}(2017)\citenamefont
  {Budrikis}, \citenamefont {Castellanos}, \citenamefont {Sandfeld},
  \citenamefont {Zaiser},\ and\ \citenamefont {Zapperi}}]{Budrikis2017}%
  \BibitemOpen
  \bibfield  {author} {\bibinfo {author} {\bibfnamefont {Z.}~\bibnamefont
  {Budrikis}}, \bibinfo {author} {\bibfnamefont {D.~F.}\ \bibnamefont
  {Castellanos}}, \bibinfo {author} {\bibfnamefont {S.}~\bibnamefont
  {Sandfeld}}, \bibinfo {author} {\bibfnamefont {M.}~\bibnamefont {Zaiser}}, \
  and\ \bibinfo {author} {\bibfnamefont {S.}~\bibnamefont {Zapperi}},\ }\href
  {\doibase 10.1038/ncomms15928} {\bibfield  {journal} {\bibinfo  {journal}
  {Nature Communications}\ }\textbf {\bibinfo {volume} {8}},\ \bibinfo {pages}
  {15928} (\bibinfo {year} {2017})}\BibitemShut {NoStop}%
\bibitem [{\citenamefont {Karimi}\ \emph {et~al.}(2017)\citenamefont {Karimi},
  \citenamefont {Ferrero},\ and\ \citenamefont {Barrat}}]{Karimi2017}%
  \BibitemOpen
  \bibfield  {author} {\bibinfo {author} {\bibfnamefont {K.}~\bibnamefont
  {Karimi}}, \bibinfo {author} {\bibfnamefont {E.~E.}\ \bibnamefont {Ferrero}},
  \ and\ \bibinfo {author} {\bibfnamefont {J.~L.}\ \bibnamefont {Barrat}},\
  }\href {\doibase 10.1103/PhysRevE.95.013003} {\bibfield  {journal} {\bibinfo
  {journal} {Physical Review E}\ }\textbf {\bibinfo {volume} {95}},\ \bibinfo
  {pages} {1} (\bibinfo {year} {2017})}\BibitemShut {NoStop}%
\bibitem [{\citenamefont {Ferrero}\ and\ \citenamefont
  {Jagla}(2019)}]{Ferrero2019}%
  \BibitemOpen
  \bibfield  {author} {\bibinfo {author} {\bibfnamefont {E.~E.}\ \bibnamefont
  {Ferrero}}\ and\ \bibinfo {author} {\bibfnamefont {E.~A.}\ \bibnamefont
  {Jagla}},\ }\href {\doibase 10.1039/c9sm01073d} {\bibfield  {journal}
  {\bibinfo  {journal} {Soft Matter}\ }\textbf {\bibinfo {volume} {15}},\
  \bibinfo {pages} {9041} (\bibinfo {year} {2019})}\BibitemShut {NoStop}%
\bibitem [{\citenamefont {Tyukodi}\ \emph {et~al.}(2019)\citenamefont
  {Tyukodi}, \citenamefont {Vandembroucq},\ and\ \citenamefont
  {Maloney}}]{Tyukodi2019}%
  \BibitemOpen
  \bibfield  {author} {\bibinfo {author} {\bibfnamefont {B.}~\bibnamefont
  {Tyukodi}}, \bibinfo {author} {\bibfnamefont {D.}~\bibnamefont
  {Vandembroucq}}, \ and\ \bibinfo {author} {\bibfnamefont {C.~E.}\
  \bibnamefont {Maloney}},\ }\href {\doibase 10.1103/PhysRevE.100.043003}
  {\bibfield  {journal} {\bibinfo  {journal} {Physical Review E}\ }\textbf
  {\bibinfo {volume} {100}},\ \bibinfo {pages} {43003} (\bibinfo {year}
  {2019})},\ \Eprint {http://arxiv.org/abs/1905.07388} {1905.07388}
  \BibitemShut {NoStop}%
\bibitem [{\citenamefont {Dahmen}\ \emph {et~al.}(2011)\citenamefont {Dahmen},
  \citenamefont {Ben-Zion},\ and\ \citenamefont {Uhl}}]{Dahmen2011}%
  \BibitemOpen
  \bibfield  {author} {\bibinfo {author} {\bibfnamefont {K.~A.}\ \bibnamefont
  {Dahmen}}, \bibinfo {author} {\bibfnamefont {Y.}~\bibnamefont {Ben-Zion}}, \
  and\ \bibinfo {author} {\bibfnamefont {J.~T.}\ \bibnamefont {Uhl}},\ }\href
  {\doibase 10.1038/nphys1957} {\bibfield  {journal} {\bibinfo  {journal}
  {Nature Physics}\ }\textbf {\bibinfo {volume} {7}},\ \bibinfo {pages} {554}
  (\bibinfo {year} {2011})}\BibitemShut {NoStop}%
\bibitem [{\citenamefont {Salje}\ and\ \citenamefont
  {Dahmen}(2014)}]{Salje2014}%
  \BibitemOpen
  \bibfield  {author} {\bibinfo {author} {\bibfnamefont {E.~K.}\ \bibnamefont
  {Salje}}\ and\ \bibinfo {author} {\bibfnamefont {K.~A.}\ \bibnamefont
  {Dahmen}},\ }\href {\doibase 10.1146/annurev-conmatphys-031113-133838}
  {\bibfield  {journal} {\bibinfo  {journal} {Annual Review of Condensed Matter
  Physics}\ }\textbf {\bibinfo {volume} {5}},\ \bibinfo {pages} {233} (\bibinfo
  {year} {2014})}\BibitemShut {NoStop}%
\bibitem [{\citenamefont {Martys}\ \emph
  {et~al.}(1991{\natexlab{b}})\citenamefont {Martys}, \citenamefont {Robbins},\
  and\ \citenamefont {Cieplak}}]{Martys1991a}%
  \BibitemOpen
  \bibfield  {author} {\bibinfo {author} {\bibfnamefont {N.}~\bibnamefont
  {Martys}}, \bibinfo {author} {\bibfnamefont {M.~O.}\ \bibnamefont {Robbins}},
  \ and\ \bibinfo {author} {\bibfnamefont {M.}~\bibnamefont {Cieplak}},\ }\href
  {\doibase 10.1103/PhysRevB.44.12294} {\bibfield  {journal} {\bibinfo
  {journal} {Physical Review B}\ }\textbf {\bibinfo {volume} {44}},\ \bibinfo
  {pages} {12294} (\bibinfo {year} {1991}{\natexlab{b}})}\BibitemShut {NoStop}%
\bibitem [{\citenamefont {Clemmer}\ and\ \citenamefont
  {Robbins}(2019)}]{Clemmer2019}%
  \BibitemOpen
  \bibfield  {author} {\bibinfo {author} {\bibfnamefont {J.~T.}\ \bibnamefont
  {Clemmer}}\ and\ \bibinfo {author} {\bibfnamefont {M.~O.}\ \bibnamefont
  {Robbins}},\ }\href {\doibase 10.1103/PhysRevE.100.042121} {\bibfield
  {journal} {\bibinfo  {journal} {Physical Review E}\ }\textbf {\bibinfo
  {volume} {100}},\ \bibinfo {pages} {42121} (\bibinfo {year} {2019})},\
  \Eprint {http://arxiv.org/abs/1909.13114} {1909.13114} \BibitemShut {NoStop}%
\bibitem [{\citenamefont {Falk}\ and\ \citenamefont {Langer}(1998)}]{Falk1997}%
  \BibitemOpen
  \bibfield  {author} {\bibinfo {author} {\bibfnamefont {M.~L.}\ \bibnamefont
  {Falk}}\ and\ \bibinfo {author} {\bibfnamefont {J.~S.}\ \bibnamefont
  {Langer}},\ }\href {\doibase 10.1103/PhysRevE.57.7192} {\bibfield  {journal}
  {\bibinfo  {journal} {Physical Review E}\ }\textbf {\bibinfo {volume} {57}},\
  \bibinfo {pages} {7192} (\bibinfo {year} {1998})},\ \Eprint
  {http://arxiv.org/abs/9712114} {9712114 [cond-mat]} \BibitemShut {NoStop}%
\bibitem [{\citenamefont {Langer}(2001)}]{Langer2001}%
  \BibitemOpen
  \bibfield  {author} {\bibinfo {author} {\bibfnamefont {J.~S.}\ \bibnamefont
  {Langer}},\ }\href {\doibase 10.1103/PhysRevE.64.011504} {\bibfield
  {journal} {\bibinfo  {journal} {Physical Review E}\ }\textbf {\bibinfo
  {volume} {64}},\ \bibinfo {pages} {011504} (\bibinfo {year}
  {2001})}\BibitemShut {NoStop}%
\bibitem [{\citenamefont {Middleton}(1992)}]{Middleton1992}%
  \BibitemOpen
  \bibfield  {author} {\bibinfo {author} {\bibfnamefont {A.~A.}\ \bibnamefont
  {Middleton}},\ }\href {\doibase 10.1103/PhysRevLett.68.670} {\bibfield
  {journal} {\bibinfo  {journal} {Physical Review Letters}\ }\textbf {\bibinfo
  {volume} {68}},\ \bibinfo {pages} {670} (\bibinfo {year} {1992})}\BibitemShut
  {NoStop}%
\bibitem [{\citenamefont {Middleton}\ and\ \citenamefont
  {Fisher}(1993)}]{Middleton1993}%
  \BibitemOpen
  \bibfield  {author} {\bibinfo {author} {\bibfnamefont {A.~A.}\ \bibnamefont
  {Middleton}}\ and\ \bibinfo {author} {\bibfnamefont {D.~S.}\ \bibnamefont
  {Fisher}},\ }\href {\doibase 10.1103/PhysRevB.47.3530} {\bibfield  {journal}
  {\bibinfo  {journal} {Physical Review B}\ }\textbf {\bibinfo {volume} {47}},\
  \bibinfo {pages} {3530} (\bibinfo {year} {1993})}\BibitemShut {NoStop}%
\bibitem [{\citenamefont {Maloney}\ and\ \citenamefont
  {Lema{\^{i}}tre}(2006)}]{Maloney2006}%
  \BibitemOpen
  \bibfield  {author} {\bibinfo {author} {\bibfnamefont {C.~E.}\ \bibnamefont
  {Maloney}}\ and\ \bibinfo {author} {\bibfnamefont {A.}~\bibnamefont
  {Lema{\^{i}}tre}},\ }\href {\doibase 10.1103/PhysRevE.74.016118} {\bibfield
  {journal} {\bibinfo  {journal} {Physical Review E}\ }\textbf {\bibinfo
  {volume} {74}},\ \bibinfo {pages} {016118} (\bibinfo {year} {2006})},\
  \Eprint {http://arxiv.org/abs/0510677} {0510677 [cond-mat]} \BibitemShut
  {NoStop}%
\bibitem [{\citenamefont {Maloney}\ and\ \citenamefont
  {Robbins}(2008)}]{Maloney2008}%
  \BibitemOpen
  \bibfield  {author} {\bibinfo {author} {\bibfnamefont {C.~E.}\ \bibnamefont
  {Maloney}}\ and\ \bibinfo {author} {\bibfnamefont {M.~O.}\ \bibnamefont
  {Robbins}},\ }\href {\doibase 10.1088/0953-8984/20/24/244128} {\bibfield
  {journal} {\bibinfo  {journal} {Journal of Physics: Condensed Matter}\
  }\textbf {\bibinfo {volume} {20}},\ \bibinfo {pages} {244128} (\bibinfo
  {year} {2008})}\BibitemShut {NoStop}%
\bibitem [{\citenamefont {Maloney}\ and\ \citenamefont
  {Robbins}(2009)}]{Maloney2009}%
  \BibitemOpen
  \bibfield  {author} {\bibinfo {author} {\bibfnamefont {C.~E.}\ \bibnamefont
  {Maloney}}\ and\ \bibinfo {author} {\bibfnamefont {M.~O.}\ \bibnamefont
  {Robbins}},\ }\href {\doibase 10.1103/PhysRevLett.102.225502} {\bibfield
  {journal} {\bibinfo  {journal} {Physical Review Letters}\ }\textbf {\bibinfo
  {volume} {102}},\ \bibinfo {pages} {225502} (\bibinfo {year}
  {2009})}\BibitemShut {NoStop}%
\bibitem [{\citenamefont {Lan{\c{c}}on}\ and\ \citenamefont
  {Billard}(1988)}]{Lan1988}%
  \BibitemOpen
  \bibfield  {author} {\bibinfo {author} {\bibfnamefont {F.}~\bibnamefont
  {Lan{\c{c}}on}}\ and\ \bibinfo {author} {\bibfnamefont {L.}~\bibnamefont
  {Billard}},\ }\href {\doibase 10.1051/jphys:01988004902024900} {\bibfield
  {journal} {\bibinfo  {journal} {Journal de Physique}\ }\textbf {\bibinfo
  {volume} {49}},\ \bibinfo {pages} {249} (\bibinfo {year} {1988})}\BibitemShut
  {NoStop}%
\bibitem [{\citenamefont {van Meel}\ \emph {et~al.}(2009)\citenamefont {van
  Meel}, \citenamefont {Charbonneau}, \citenamefont {Fortini},\ and\
  \citenamefont {Charbonneau}}]{VanMeel2009}%
  \BibitemOpen
  \bibfield  {author} {\bibinfo {author} {\bibfnamefont {J.~A.}\ \bibnamefont
  {van Meel}}, \bibinfo {author} {\bibfnamefont {B.}~\bibnamefont
  {Charbonneau}}, \bibinfo {author} {\bibfnamefont {A.}~\bibnamefont
  {Fortini}}, \ and\ \bibinfo {author} {\bibfnamefont {P.}~\bibnamefont
  {Charbonneau}},\ }\href {\doibase 10.1103/PhysRevE.80.061110} {\bibfield
  {journal} {\bibinfo  {journal} {Physical Review E}\ }\textbf {\bibinfo
  {volume} {80}},\ \bibinfo {pages} {061110} (\bibinfo {year}
  {2009})}\BibitemShut {NoStop}%
\bibitem [{\citenamefont {Plimpton}(1995)}]{Plimpton1995}%
  \BibitemOpen
  \bibfield  {author} {\bibinfo {author} {\bibfnamefont {S.}~\bibnamefont
  {Plimpton}},\ }\href {\doibase 10.1006/jcph.1995.1039} {\bibfield  {journal}
  {\bibinfo  {journal} {Journal of Computational Physics}\ }\textbf {\bibinfo
  {volume} {117}},\ \bibinfo {pages} {1} (\bibinfo {year} {1995})}\BibitemShut
  {NoStop}%
\bibitem [{\citenamefont {Allen}\ and\ \citenamefont
  {Tildesley}(1989)}]{Allen1989}%
  \BibitemOpen
  \bibfield  {author} {\bibinfo {author} {\bibfnamefont {M.}~\bibnamefont
  {Allen}}\ and\ \bibinfo {author} {\bibfnamefont {D.}~\bibnamefont
  {Tildesley}},\ }\href {https://books.google.com/books?id=O32VXB9e5P4C} {\emph
  {\bibinfo {title} {Computer Simulation of Liquids}}},\ Oxford Science Publ\
  (\bibinfo  {publisher} {Clarendon Press},\ \bibinfo {year}
  {1989})\BibitemShut {NoStop}%
\bibitem [{\citenamefont {Kraynik}\ and\ \citenamefont
  {Reinelt}(1992)}]{Kraynik1992}%
  \BibitemOpen
  \bibfield  {author} {\bibinfo {author} {\bibfnamefont {A.}~\bibnamefont
  {Kraynik}}\ and\ \bibinfo {author} {\bibfnamefont {D.}~\bibnamefont
  {Reinelt}},\ }\href {\doibase 10.1016/0301-9322(92)90074-Q} {\bibfield
  {journal} {\bibinfo  {journal} {International Journal of Multiphase Flow}\
  }\textbf {\bibinfo {volume} {18}},\ \bibinfo {pages} {1045} (\bibinfo {year}
  {1992})}\BibitemShut {NoStop}%
\bibitem [{\citenamefont {Hunt}(2016)}]{Hunt2016}%
  \BibitemOpen
  \bibfield  {author} {\bibinfo {author} {\bibfnamefont {T.~A.}\ \bibnamefont
  {Hunt}},\ }\href {\doibase 10.1080/08927022.2015.1051043} {\bibfield
  {journal} {\bibinfo  {journal} {Molecular Simulation}\ }\textbf {\bibinfo
  {volume} {42}},\ \bibinfo {pages} {347} (\bibinfo {year} {2016})}\BibitemShut
  {NoStop}%
\bibitem [{\citenamefont {Nicholson}\ and\ \citenamefont
  {Rutledge}(2016)}]{Nicholson2016}%
  \BibitemOpen
  \bibfield  {author} {\bibinfo {author} {\bibfnamefont {D.~A.}\ \bibnamefont
  {Nicholson}}\ and\ \bibinfo {author} {\bibfnamefont {G.~C.}\ \bibnamefont
  {Rutledge}},\ }\href {\doibase 10.1063/1.4972894} {\bibfield  {journal}
  {\bibinfo  {journal} {The Journal of Chemical Physics}\ }\textbf {\bibinfo
  {volume} {145}},\ \bibinfo {pages} {244903} (\bibinfo {year}
  {2016})}\BibitemShut {NoStop}%
\bibitem [{\citenamefont {Evans}\ and\ \citenamefont
  {Morriss}(1984)}]{Evans1984}%
  \BibitemOpen
  \bibfield  {author} {\bibinfo {author} {\bibfnamefont {D.~J.}\ \bibnamefont
  {Evans}}\ and\ \bibinfo {author} {\bibfnamefont {G.~P.}\ \bibnamefont
  {Morriss}},\ }\href {\doibase 10.1103/PhysRevA.30.1528} {\bibfield  {journal}
  {\bibinfo  {journal} {Physical Review A}\ }\textbf {\bibinfo {volume} {30}},\
  \bibinfo {pages} {1528} (\bibinfo {year} {1984})}\BibitemShut {NoStop}%
\bibitem [{\citenamefont {Varnik}\ \emph {et~al.}(2004)\citenamefont {Varnik},
  \citenamefont {Bocquet},\ and\ \citenamefont {Barrat}}]{Varnik2004}%
  \BibitemOpen
  \bibfield  {author} {\bibinfo {author} {\bibfnamefont {F.}~\bibnamefont
  {Varnik}}, \bibinfo {author} {\bibfnamefont {L.}~\bibnamefont {Bocquet}}, \
  and\ \bibinfo {author} {\bibfnamefont {J.~L.}\ \bibnamefont {Barrat}},\
  }\href {\doibase 10.1063/1.1636451} {\bibfield  {journal} {\bibinfo
  {journal} {Journal of Chemical Physics}\ }\textbf {\bibinfo {volume} {120}},\
  \bibinfo {pages} {2788} (\bibinfo {year} {2004})}\BibitemShut {NoStop}%
\bibitem [{\citenamefont {Shi}\ and\ \citenamefont {Falk}(2005)}]{Shi2005}%
  \BibitemOpen
  \bibfield  {author} {\bibinfo {author} {\bibfnamefont {Y.}~\bibnamefont
  {Shi}}\ and\ \bibinfo {author} {\bibfnamefont {M.~L.}\ \bibnamefont {Falk}},\
  }\href {\doibase 10.1103/PhysRevLett.95.095502} {\bibfield  {journal}
  {\bibinfo  {journal} {Physical Review Letters}\ }\textbf {\bibinfo {volume}
  {95}},\ \bibinfo {pages} {095502} (\bibinfo {year} {2005})},\ \Eprint
  {http://arxiv.org/abs/0503285} {0503285 [cond-mat]} \BibitemShut {NoStop}%
\bibitem [{\citenamefont {Rottler}\ and\ \citenamefont
  {Robbins}(2005)}]{Rottler2005}%
  \BibitemOpen
  \bibfield  {author} {\bibinfo {author} {\bibfnamefont {J.}~\bibnamefont
  {Rottler}}\ and\ \bibinfo {author} {\bibfnamefont {M.~O.}\ \bibnamefont
  {Robbins}},\ }\href {\doibase 10.1103/PhysRevLett.95.225504} {\bibfield
  {journal} {\bibinfo  {journal} {Physical Review Letters}\ }\textbf {\bibinfo
  {volume} {95}},\ \bibinfo {pages} {225504} (\bibinfo {year}
  {2005})}\BibitemShut {NoStop}%
\bibitem [{\citenamefont {Ozawa}\ \emph {et~al.}(2018)\citenamefont {Ozawa},
  \citenamefont {Berthier}, \citenamefont {Biroli}, \citenamefont {Rosso},\
  and\ \citenamefont {Tarjus}}]{Ozawa2018}%
  \BibitemOpen
  \bibfield  {author} {\bibinfo {author} {\bibfnamefont {M.}~\bibnamefont
  {Ozawa}}, \bibinfo {author} {\bibfnamefont {L.}~\bibnamefont {Berthier}},
  \bibinfo {author} {\bibfnamefont {G.}~\bibnamefont {Biroli}}, \bibinfo
  {author} {\bibfnamefont {A.}~\bibnamefont {Rosso}}, \ and\ \bibinfo {author}
  {\bibfnamefont {G.}~\bibnamefont {Tarjus}},\ }\href {\doibase
  10.1073/pnas.1806156115} {\bibfield  {journal} {\bibinfo  {journal}
  {Proceedings of the National Academy of Sciences}\ }\textbf {\bibinfo
  {volume} {115}},\ \bibinfo {pages} {6656} (\bibinfo {year}
  {2018})}\BibitemShut {NoStop}%
\bibitem [{\citenamefont {Clemmer}\ \emph {et~al.}(2021)\citenamefont
  {Clemmer}, \citenamefont {Salerno},\ and\ \citenamefont
  {Robbins}}]{Clemmer2021}%
  \BibitemOpen
  \bibfield  {author} {\bibinfo {author} {\bibfnamefont {J.~T.}\ \bibnamefont
  {Clemmer}}, \bibinfo {author} {\bibfnamefont {K.~M.}\ \bibnamefont
  {Salerno}}, \ and\ \bibinfo {author} {\bibfnamefont {M.~O.}\ \bibnamefont
  {Robbins}},\ }\href@noop {} {\  (\bibinfo {year} {2021})}\BibitemShut
  {NoStop}%
\bibitem [{\citenamefont {Maloney}(2015)}]{Maloney2015}%
  \BibitemOpen
  \bibfield  {author} {\bibinfo {author} {\bibfnamefont {C.~E.}\ \bibnamefont
  {Maloney}},\ }\href {\doibase 10.1209/0295-5075/111/28001} {\bibfield
  {journal} {\bibinfo  {journal} {EPL (Europhysics Letters)}\ }\textbf
  {\bibinfo {volume} {111}},\ \bibinfo {pages} {28001} (\bibinfo {year}
  {2015})}\BibitemShut {NoStop}%
\bibitem [{Note1()}]{Note1}%
  \BibitemOpen
  \bibinfo {note} {Note however that snapshots of $\sigma $ over short strain
  intervals do not have a clear trend with rate because of the large
  fluctuations in the instantaneous shear stress discussed in Sec. \ref
  {sec:yield_deviate}}\BibitemShut {NoStop}%
\bibitem [{\citenamefont {Lema{\^{i}}tre}\ and\ \citenamefont
  {Caroli}(2007)}]{Lemaitre2007}%
  \BibitemOpen
  \bibfield  {author} {\bibinfo {author} {\bibfnamefont {A.}~\bibnamefont
  {Lema{\^{i}}tre}}\ and\ \bibinfo {author} {\bibfnamefont {C.}~\bibnamefont
  {Caroli}},\ }\href {\doibase 10.1103/PhysRevE.76.036104} {\bibfield
  {journal} {\bibinfo  {journal} {Physical Review E}\ }\textbf {\bibinfo
  {volume} {76}},\ \bibinfo {pages} {036104} (\bibinfo {year}
  {2007})}\BibitemShut {NoStop}%
\bibitem [{\citenamefont {Salerno}(2013)}]{SalernoThesis}%
  \BibitemOpen
  \bibfield  {author} {\bibinfo {author} {\bibfnamefont {K.~M.}\ \bibnamefont
  {Salerno}},\ }\emph {\bibinfo {title} {Inertia and the Critical Scaling of
  Avalanches in Sheared Disordered Solids}},\ \href@noop {} {Ph.D. thesis},\
  \bibinfo  {school} {Johns Hopkins University} (\bibinfo {year}
  {2013})\BibitemShut {NoStop}%
\bibitem [{\citenamefont {Tyukodi}\ \emph {et~al.}(2018)\citenamefont
  {Tyukodi}, \citenamefont {Vandembroucq},\ and\ \citenamefont
  {Maloney}}]{Tyukodi2018}%
  \BibitemOpen
  \bibfield  {author} {\bibinfo {author} {\bibfnamefont {B.}~\bibnamefont
  {Tyukodi}}, \bibinfo {author} {\bibfnamefont {D.}~\bibnamefont
  {Vandembroucq}}, \ and\ \bibinfo {author} {\bibfnamefont {C.~E.}\
  \bibnamefont {Maloney}},\ }\href {\doibase 10.1103/PhysRevLett.121.145501}
  {\bibfield  {journal} {\bibinfo  {journal} {Physical Review Letters}\
  }\textbf {\bibinfo {volume} {121}},\ \bibinfo {pages} {145501} (\bibinfo
  {year} {2018})},\ \Eprint {http://arxiv.org/abs/1803.06009} {1803.06009}
  \BibitemShut {NoStop}%
\bibitem [{\citenamefont {Lema{\^{i}}tre}\ and\ \citenamefont
  {Caroli}(2009)}]{Lemaitre2009}%
  \BibitemOpen
  \bibfield  {author} {\bibinfo {author} {\bibfnamefont {A.}~\bibnamefont
  {Lema{\^{i}}tre}}\ and\ \bibinfo {author} {\bibfnamefont {C.}~\bibnamefont
  {Caroli}},\ }\href {\doibase 10.1103/PhysRevLett.103.065501} {\bibfield
  {journal} {\bibinfo  {journal} {Physical Review Letters}\ }\textbf {\bibinfo
  {volume} {103}},\ \bibinfo {pages} {065501} (\bibinfo {year}
  {2009})}\BibitemShut {NoStop}%
\bibitem [{\citenamefont {Lin}\ and\ \citenamefont {Wyart}(2018)}]{Lin2018}%
  \BibitemOpen
  \bibfield  {author} {\bibinfo {author} {\bibfnamefont {J.}~\bibnamefont
  {Lin}}\ and\ \bibinfo {author} {\bibfnamefont {M.}~\bibnamefont {Wyart}},\
  }\href {\doibase 10.1103/PhysRevE.97.012603} {\bibfield  {journal} {\bibinfo
  {journal} {Physical Review E}\ }\textbf {\bibinfo {volume} {97}},\ \bibinfo
  {pages} {012603} (\bibinfo {year} {2018})}\BibitemShut {NoStop}%
\bibitem [{\citenamefont {Chaudhuri}\ \emph {et~al.}(2012)\citenamefont
  {Chaudhuri}, \citenamefont {Berthier},\ and\ \citenamefont
  {Bocquet}}]{Chaudhuri2012}%
  \BibitemOpen
  \bibfield  {author} {\bibinfo {author} {\bibfnamefont {P.}~\bibnamefont
  {Chaudhuri}}, \bibinfo {author} {\bibfnamefont {L.}~\bibnamefont {Berthier}},
  \ and\ \bibinfo {author} {\bibfnamefont {L.}~\bibnamefont {Bocquet}},\ }\href
  {\doibase 10.1103/PhysRevE.85.021503} {\bibfield  {journal} {\bibinfo
  {journal} {Physical Review E}\ }\textbf {\bibinfo {volume} {85}},\ \bibinfo
  {pages} {021503} (\bibinfo {year} {2012})},\ \Eprint
  {http://arxiv.org/abs/1111.5957} {1111.5957} \BibitemShut {NoStop}%
\bibitem [{\citenamefont {Karmakar}\ \emph {et~al.}(2010)\citenamefont
  {Karmakar}, \citenamefont {Lerner}, \citenamefont {Procaccia},\ and\
  \citenamefont {Zylberg}}]{Karmakar2010a}%
  \BibitemOpen
  \bibfield  {author} {\bibinfo {author} {\bibfnamefont {S.}~\bibnamefont
  {Karmakar}}, \bibinfo {author} {\bibfnamefont {E.}~\bibnamefont {Lerner}},
  \bibinfo {author} {\bibfnamefont {I.}~\bibnamefont {Procaccia}}, \ and\
  \bibinfo {author} {\bibfnamefont {J.}~\bibnamefont {Zylberg}},\ }\href
  {\doibase 10.1103/PhysRevE.82.031301} {\bibfield  {journal} {\bibinfo
  {journal} {Physical Review E}\ }\textbf {\bibinfo {volume} {82}},\ \bibinfo
  {pages} {031301} (\bibinfo {year} {2010})},\ \Eprint
  {http://arxiv.org/abs/1006.3737} {1006.3737} \BibitemShut {NoStop}%
\bibitem [{\citenamefont {P{\'{a}}zm{\'{a}}ndi}\ \emph
  {et~al.}(1997)\citenamefont {P{\'{a}}zm{\'{a}}ndi}, \citenamefont
  {Scalettar},\ and\ \citenamefont {Zim{\'{a}}nyi}}]{Pazmandi1997}%
  \BibitemOpen
  \bibfield  {author} {\bibinfo {author} {\bibfnamefont {F.}~\bibnamefont
  {P{\'{a}}zm{\'{a}}ndi}}, \bibinfo {author} {\bibfnamefont {R.~T.}\
  \bibnamefont {Scalettar}}, \ and\ \bibinfo {author} {\bibfnamefont {G.~T.}\
  \bibnamefont {Zim{\'{a}}nyi}},\ }\href {\doibase 10.1103/PhysRevLett.79.5130}
  {\bibfield  {journal} {\bibinfo  {journal} {Physical Review Letters}\
  }\textbf {\bibinfo {volume} {79}},\ \bibinfo {pages} {5130} (\bibinfo {year}
  {1997})}\BibitemShut {NoStop}%
\end{thebibliography}

\newpage

\end{document}